\documentclass[aps,reprint,showpacs,superscriptaddress]{revtex4-2}

\usepackage{amssymb}
\usepackage{natbib}
\usepackage{graphicx}
\usepackage{amsmath}
\usepackage[bookmarks = false]{hyperref}
\usepackage{bm}
\usepackage[all]{hypcap}
\usepackage{colortbl}
\usepackage{booktabs}
\usepackage{braket}
\usepackage{mathrsfs}
\usepackage{multirow}
\usepackage{upgreek}
\usepackage{textcomp}
\usepackage{soul}
\usepackage{dsfont}
\usepackage[T1]{fontenc}

\usepackage{lineno}

\usepackage{array}
\newcommand{\PreserveBackslash}[1]{\let\temp=\\#1\let\\=\temp}
\newcolumntype{C}[1]{>{\PreserveBackslash\centering}p{#1}}
\newcolumntype{R}[1]{>{\PreserveBackslash\raggedleft}p{#1}}
\newcolumntype{L}[1]{>{\PreserveBackslash\raggedright}p{#1}}

\newcommand{\ustc}{
	\affiliation{Hefei National Laboratory for Physical Sciences at Microscale and Department
	of Modern Physics, University of Science and Technology of China, Hefei,
	Anhui 230026, China}
	\affiliation{CAS Center for Excellence in Quantum Information and Quantum Physics, University of Science and Technology of China, Hefei, Anhui 230026, China}
	\affiliation{Hefei National Laboratory, University of Science and Technology of China, Hefei 230088, China}
}

\newcommand{\hfnl}{
	\affiliation{Hefei National Laboratory, University of Science and Technology of China, Hefei 230088, China}
}

\newcommand{\jinan}{
	\affiliation{Jinan Institute of Quantum Technology, Jinan, China}
}

\newcommand{\snspd}{
	\affiliation{Shanghai Key Laboratory of Superconductor Integrated Circuit Technology, Shanghai Institute of Microsystem and Information Technology, Chinese Academy of Sciences, Shanghai 200050, China}
}
\newcommand{\fiber}{
	\affiliation{State Key Laboratory of Optical Fibre and Cable Manufacture Technology, Yangtze Optical Fibre and Cable Joint Stock Limited Company, Wuhan 430073, China}
}

\begin{document}

\title{Entangling quantum memories over 420 km in fiber}

\author{Xi-Yu Luo}\ustc
\author{Chao-Yang Wang}\ustc
\author{Ming-Yang Zheng}\jinan\hfnl
\author{Bin Wang}\ustc
\author{Jian-Long Liu}\ustc
\author{Bo-Feng Gao}\jinan
\author{Jun Li}\ustc
\author{Zi Yan}\ustc
\author{Qiao-Mu Ke}\ustc
\author{Da Teng}\ustc
\author{Rui-Chun Wang}\fiber
\author{Jun Wu}\fiber
\author{Jia Huang}\snspd
\author{Hao Li}\snspd
\author{Li-Xing You}\snspd
\author{Xiu-Ping Xie}\jinan\hfnl
\author{Feihu Xu}\ustc
\author{Qiang Zhang}\ustc\jinan
\author{Xiao-Hui Bao}\ustc
\author{Jian-Wei Pan}\ustc

\begin{abstract}
	\normalsize
	Long-distance entanglement is pivotal for quantum communication~\cite{kimble_internet_2008,wehner_quantum_2018}, distributed quantum computing~\cite{jiang_distributedcomputation_2007} and sensing~\cite{gottesman_sensing_2012,komar_clock_2014}. Significant progresses have been made in extending the distribution distance of entangled photons, either in free space~\cite{lu_micius_2022} or fiber~\cite{inagaki_entanglement_2013,neumann_entanglement_2022,zhuang_ultrabright-entanglement-based_2024}. For future quantum network applications~\cite{wehner_quantum_2018}, matter-based entanglement is more favorable since the capability of storage is essential for advanced applications~\cite{azuma_quantum_2023}. Extending entanglement distance for memory qubits was partially hindered by the mismatch of its photonic emission wavelength with the low-loss transmission window of optical fiber. By incorporating quantum frequency conversion~\cite{kumar_quantum_1990}, memory-memory entanglement has been successfully extended to several tens of kilometers~\cite{yu_entanglement_2020,van_leent_entangling_2022,knaut_entanglement_2024,stolk_metropolitan-scale_2024,liu_creation_2024}. Here, we make a significant step further by reporting the entanglement between two atomic ensemble quantum memories over 420 km. We convert photons emitted from the memories to telecom S-band, which enable us to exploit the significantly low transmission loss in fiber (0.17 dB/km). We employ the DLCZ scheme~\cite{duan_long-distance_2001} for remote entanglement generation, and delicately stabilize the relative phase between the two memories by using fulltime far-off-resonant locking to reduce high-frequency noise and intermittent dual-band locking to compensate low-frequency drift jointly. We demonstrate that the memory-memory entangling probability beats the repeaterless channel capacity~\cite{pirandola_fundamental_2017} for direct entanglement distribution. Our experiment provides a testbed of studying quantum network applications from metropolitan scale to intercity scale.
\end{abstract}

\maketitle

Long-distance distribution of entangled photons has been a great success in recent years. By sending entangled photons from a satellite~\cite{lu_micius_2022}, entanglement was measured at ground stations separated over 1200 km~\cite{yin_satellite-based_2017}, leveraging low-loss transmission in the outside of atmosphere. In the fiber link, long-distance entangled photon distribution is difficult, due to the exponential decay of transmittance as the fiber goes longer. A very recent result demonstrates the fiber-based entanglement distribution up to 404 km~\cite{zhuang_ultrabright-entanglement-based_2024}. In these experiments, most of the photons are lost during travel to the detectors, and the overall success probability is rather low. For advanced quantum network applications~\cite{wehner_quantum_2018}, it is much favorable that the success of entangled photon distribution is heralded and stored for latter use. 

\begin{figure*}[tp]
	\centering
	\includegraphics[width=.8\textwidth]{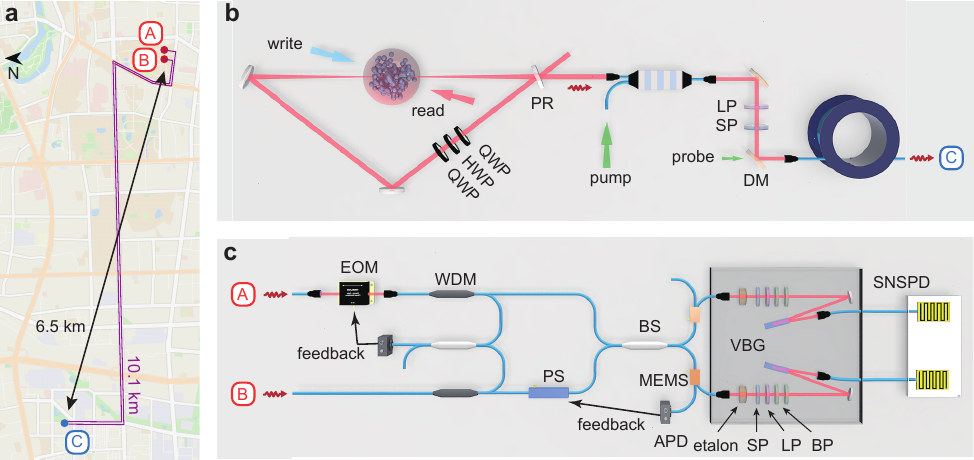}
	 \caption{\textbf{Experimental layout.} 
	 \textbf{a,} Bird's eye view of our experiment setup. Two memory nodes Alice and Bob are all located at a laboratory in USTC. A pair of deployed fibers of 10.1~km are used to transmit write-out photons from two nodes to the middle node Charlie at Hefei Software Park. \textbf{b,} Each memory node includes a cavity-enhanced DLCZ-type memory and a QFC module. The write-out photon is shifted to the telecom S-band with a QFC module and sent to node Charlie through some coiled fibers and the 10.1~km field fiber. Meanwhile, a 1600~nm phase probe light is coupled to the same path through the second DM. PR: partially reflective mirror, DM: dichroic mirror, LP(SP): long(short)-pass filter, H(Q)WP: half(quarter)-wave-plate. \textbf{c,} At node Charlie, the probe light from two nodes can be filtered out by wavelength division multiplexer (WDM) and interfered at a beamsplitter (BS). The interference signal is collected by avalanche photon diode (APD) and fed back to the EOM. The 1522~nm signal photons are then interfered at another BS. A combination of etalon, LP, BP are used to filter residual Raman noise in the conversion and transmission process. Finally, the photons are detected by two superconducting nanowire single-photon detectors (SNSPDs). The MEMSs are switched to APD for phase locking during the memory preparation phase. PS: phase shifter. Map data from Mapbox and OpenStreetMap.}
	\label{fig:setup}
\end{figure*}

In the context of quantum networks, a main focus is to entangle matter qubits over long distance. A typical way is to create atom-photon entanglement in two end nodes, and send the photons to a middle node for interference and detection. The detection of photons heralds the success of atom-atom entanglement between the two end nodes, and the atoms act as quantum memories. Along this direction, many different physical systems have been investigated~\cite{azuma_quantum_2023}, such as laser-cooled atomic ensembles~\cite{chou_measurement-induced_2005,chou_entanglement_2007,yuan_bdcz_2008}, solid-state rare earth ions~\cite{lago-rivera_rareearth_2021,liu_rareearth_2021}, quantum dots~\cite{delteil_quantumdot_2016,stockill_quantumdot_2017}, as well as color centers~\cite{bernien_colorcenter_2013,hensen_colorcenter_2015,humphreys_colorcenter_2018}, trapped single ions~\cite{moehring_ions_2007} and single atoms~\cite{hofmann_singleatom_2012}. In extending the entangling distance of these matter-based qubits, a common problem is the mismatch of wavelength. For optical fibers, the lowest transmission losses occur in the wavelength range of 1300 nm to 1600 nm. While, the photon emission of matter qubits typically occurs in the range of visible or near infrared. To mitigate this mismatch, the technology of quantum frequency conversion (QFC)~\cite{kumar_quantum_1990} was employed to convert photons emitted from matter qubits to a wavelength in the telecom range. With this technology, previous experiments have demonstrated entanglement of matter qubits over several tens of kilometers~\cite{yu_entanglement_2020,van_leent_entangling_2022,knaut_entanglement_2024,stolk_metropolitan-scale_2024}, and matter-photon entanglement over 100~km~\cite{zhou_atomphotonentanglement_2024,krutyanskiy_ionphotonentanglement_2024} as well. Meanwhile, a metropolitan-scale multinode quantum network~\cite{liu_creation_2024} has been established. Entanglement of matter qubits at larger scale, for instance to enable inter-city quantum networking, is still lacking. 

In this paper, we report the generation of atom-atom entanglement over fiber transmission of 420 km. We make use of laser-cooled atomic ensembles and harness the Duan-Lukin-Cirac-Zoller (DLCZ) scheme~\cite{duan_long-distance_2001} for entanglement generation. Via interfering and detecting merely a single photon, the success probability of remote entanglement generation is significantly higher than two-photon based schemes~\cite{simon_entanglement_2003,zhao_entanglement_2007} and atom-photon-gate based schemes~\cite{severin_gate_2021,knaut_entanglement_2024}. To mitigate phase fluctuations during long fiber transmission, we have developed an exquisite phase stabilization scheme which combines full-time stabilization to reduce high-frequency noise and intermittent stabilization to compensate low-frequency drift. With this scheme, the overall phase stability is around 7$^\circ$. The created atom-atom entanglement is verified via retrieving the atomic state back to photons and let them interfere. We perform measurement for a series of different fiber lengths. Up to 420 km, atom-atom entanglement is genuinely verified. In comparison with direct transmission of perfect entanglement photons in fibers, our experiment has a much higher success probability. Harnessed with storage capability and potential of deterministic atomic state detection~\cite{yang_deterministic_2022}, our experiment may find much broader quantum network applications beyond entangled photons, such as device-independent quantum key distribution (DI-QKD)~\cite{zhang_device-independent_2022,nadlinger_experimental_2022}. 

\begin{figure*}[htbp]
	\centering
	\includegraphics[width=\textwidth]{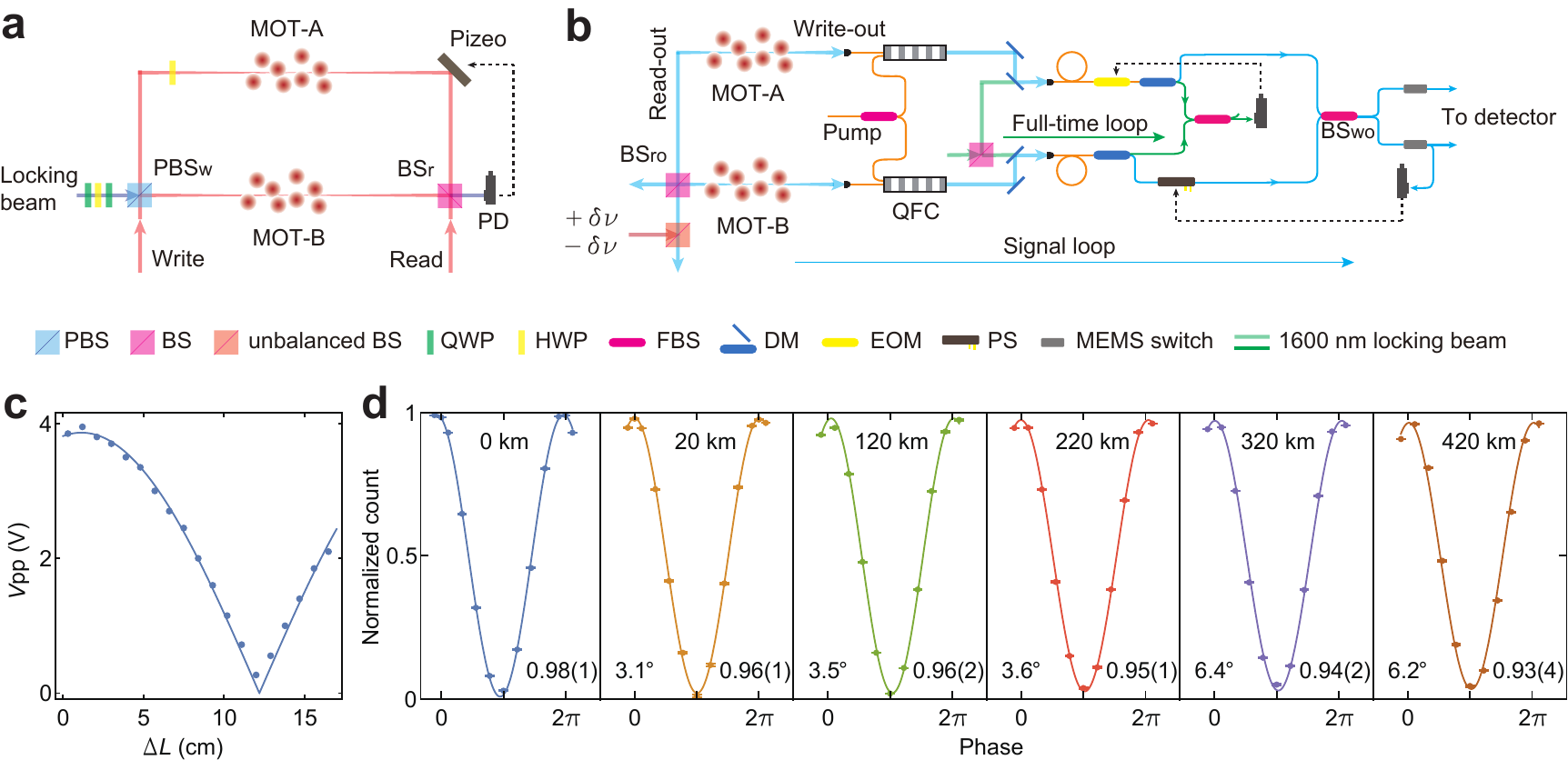}
	\caption{\textbf{Phase stabilization between two nodes.}  
	Phase stabilization includes two individual interferometers. \textbf{a,}  Configuration of the local write-read interferometer. This interferometer covers paths of write and read beams. We introduce a relative locking phase $\theta$ between two arms by a sandwich configuration of wave plates. \textbf{b,} Configuration of the remote write-out-read-out interferometer. This interferometer covers paths of write-out and read-out photons including the QFC modules. \textbf{c,}  Interference peak-to-peak value of the dual-band locking beam as a function of $\Delta L$. The curve is a sinusoidal fitting of the data points. \textbf{d,} Characterization of the remote phase stabilization. A weak probe pulse is lead in from the BS$_{\rm ro}$ and interfere at the BS$_{\rm wo}$. The data points show the photon count in one output mode of the BS$_{\rm wo}$ as a function of the locking phase difference between two arms, normalized by the total count of the maximum and minimum of the fitted curve. The fitted visibility of each curve and measured residual phase error is shown at the bottom right (left). Error bars indicate one standard deviation of the photon-counting statistics.
	}
	\label{fig:phase detail}
\end{figure*}

\section*{Scheme and experimental setup}

As shown in Fig.~\ref{fig:setup}, in each memory node, all atoms are initially prepared in the ground state $\ket{g}=\ket{5S_{1/2}, F=1, m_F=1}$. A weak 780~nm write pulse blue detuned coupling it to the excitation state $\ket{e}=\ket{5P_{3/2}, F=2, m_F=0}$ induces a spontaneous Raman-scattered (write out) photon and a collective excitation in a stable state $\ket{s}=\ket{5S_{1/2}, F=2, m_F=1}$ with a small probability ($\chi\approx 6\%$). The ensemble and the write-out photon form a Fock state entanglement (without normalization): $\ket{\Psi_{ap}}=\ket{0_a0_p}+\sqrt{\chi}\ket{1_a1_p}$, where 0 and 1 refer to the number of atomic (subscript a) or photon (subscript p)  excitations. Then we shift the wavelength of the write-out photon from 780~nm to 1522~nm (telecom S-band) via the differential-frequency generation (DFG) process using a periodically poled lithium niobate waveguide (PPLN-WG) chip and a strong 1600~nm pump laser, which reduces the photon attenuation from 3.5~dB/km to 0.18~dB/km under transmission through standard optical fiber (G.652, the actual attenuation of the field fiber between nodes Alice (Bob) and Charlie is 0.31~dB/km, while the pure attenuation (without flange loss) of the ultra-low-loss coiled fiber (G.654.E) we use to extend the experiment distance is slightly lower than 0.17~dB/km at 1522~nm). After frequency conversion and broadband filtering, the entangled photons are then transmitted along a long coiled fiber and a field-deployed fiber of 10.1~km to the node Charlie at Hefei Software Park. At Charlie, they are combined with a beam splitter (BS$_{\rm wo}$) to erase ``which way'' information. Then, the photons are transmitted through narrow-band filter modules and detected by SNSPDs. A click from the detector maps two ensembles into a maximal entanglement state
\begin{equation}\label{eq:EntFock}
		\ket{\Psi^{\pm}}=(\ket{0_A1_B}\pm e^{i\Delta\varphi}\ket{1_A0_B})/\sqrt{2},
\end{equation}
where $\Delta\varphi$ is the accumulated phase difference during the entanglement process, which must be locked to maintain the coherence of the entanglement state. The collective excitation of two memories can be stored for a long time~\cite{yang_efficient_2016, wang_cavity-enhanced_2021} and retrieved to verify the entanglement on demand. Noise level and phase stabilization are two key challenges in long-distance experiments.

The noise mainly comes from three parts: the QFC process, long fibers and SNSPDs. At the QFC process, the wavelength of the pump laser is closer to the signal photon than before~\cite{yu_entanglement_2020,luo_postselected_2022,liu_creation_2024}, which produces wide-band anti-Stokes Raman (ASR) noise at the signal wavelength in the conversion waveguide. To reduce the noise level~\cite{zaske_efficient_2011,van_leent_long-distance_2020,van_leent_entangling_2022} while maintaining the low-loss advantage, we have selected 1522~nm as the signal wavelength, which is separated from the pump wavelength by 78~nm and lies marginally below the telecom C-band.

During the long fiber transmission process, the 1600~nm probe light used to lock the phase also generates Raman noise at 1522~nm, which must be filtered out before detection. So as shown in Fig.~\ref{fig:setup}, we design two modules to suppress all noises described above. At node Alice (Bob), a broadband filter module, which includes two dichroic mirrors (DM) and a pair of long (LP) and short (SP) pass filters (see supplementary for detail), is placed just behind the waveguide to filter the pump laser and the noise from its second and third harmonic generation. At node Charlie, a narrow-band filter module which mainly includes a 70~MHz bandwidth etalon, a 19~GHz bandwidth volume Bragg grating (VBG) , and several LP, SP, bandpass filters (BP), is placed just before the SNSPDs to further reduce other noises. The whole narrow-band modules are stored inside an insulated plastic container to reduce temperature disturbance while we utilize thermoelectric cooler technology to regulate the temperature of the etalon, enabling precise adjustment and stabilization of its length. To test the end-to-end performance of our QFC module, we directly connect the two modules locally and achieve an efficiency of about 44\% while the noise can be confined to about 3.2~kHz, which corresponds to a signal-to-noise ratio (SNR) of >40:1(see supplementary for details). 

In the long-distance experiment, the signal photon needs to transmit hundreds of kilometres of fiber to interfere and the remote click rate can still drop to Hertz level though we use the single-photon scheme. So two low-dark-count (about 1~Hz) detectors are used in our experiments. Their efficiencies is about 60\% ($D_1$) and 30\% ($D_2$) respectively. All SNR results of different distance experiments are listed in Tab.~\ref{tab:concurrence}.

Node Alice and Bob share three lasers including two lasers of 780~nm for manipulating atoms and one laser of 1600~nm for QFC pumping. They are all locked onto a reference cavity. Combining with the entanglement verification method, we establish two interferometers to stabilize the measurement phase: the write-read interferometer and the write-out-read-out interferometer. The former is shown in Fig.~\ref{fig:phase detail}a. The write (read) pulse is split by a PBS$_{w}$(BS$_{r}$) and injected into atomic ensembles respectively. We induce a locking beam from another input of the PBS$_{w}$ and stabilize the phase between two channels by using a piezoelectric ceramic. 

The write-out-read-out interferometer incorporates long fiber sections whose stability is crucial to the success of the entire experiment. This setup presents several key challenges. Intermittent locking fails to suppress high-frequency noise in long fiber loop. Since  the optical path passes through the atoms, locking beam with a distinct frequency is required to prevent photon-atom interaction, which inevitably induces low-frequency drift at the locking point. To address these challenges, as illustrated in Fig.~\ref{fig:phase detail}b, we developed two interferometer loops and implemented a more advanced phase-locking strategy.

To mitigate the high-frequency noise in the long fiber, we designed a full-time loop. As illustrated in Fig.~\ref{fig:phase detail}b, a 1600~nm locking beam, which is derived from our pump laser, is lead in right behind the QFC modules at node Alice (Bob), and lead out before BS$_{\rm wo}$ at node Charlie with the help of DM mirror and wavelength division multiplexer (WDM) respectively. The locking beam then interferes at another BS and we detect the interference signal and feed it back to an electro-optical modulator (EOM). Thanks to the substantial wavelength difference, the stabilization process could run continuously and the noise introduced by the probe light can be effectively filtered, keeping its contribution below 1 Hz throughout the experiment.

To address the residual path phase, we introduced an additional locking beam derived from the BS at read-out interference path (BS$_{\rm ro}$) using an unbalanced BS (T:R~$\approx$~88:12) to stabilize the phase of the entire signal loop during the MOT preparation phase. The beam is lead out after BS$_{\rm wo}$ at the node Charlie by using an optic switch based on the micro-electro-mechanical system (MEMS) technique. We feed the interference signal back to a piezoelectric stretcher and hold the voltage at the entanglement generation phase. In the former experiment~\cite{yu_entanglement_2020}, this locking beam is far detuned from the resonance point of the ring cavity and also the frequency of the write-out photon to avoid any interaction with atoms. But as the fiber length increases to several hundred kilometres, this detuning can bring in considerable phase locking uncertainty caused by the slow drift of the length difference of two write-out photon interference paths though all fiber spools are also stored inside some insulated plastic boxes. To mitigate this issue, the locking beam is replaced by a beam with two frequency components (dual-band). The detunings of two components are symmetric around the frequency of the write-out photon ($\delta\nu=675$~MHz). After a detailed derivation, the final interference intensity can be expressed as: 
\begin{equation}\label{eq:IntSig}
I\approx S+\tilde{A} \cos k_0\Delta L,
\end{equation}
where $S$ represents a constant offset, $\Delta L$ denotes the length difference between the two interferometer arms and $k_0$ corresponds to the wave number of the write-out photon rather than the detuned locking beam. The term $\tilde{A}$ ($\varpropto\cos\delta k\Delta L$, where $\delta k=k_+-k_0\ll k_0$ and $k_+$ is the wave number of the blue-detuned locking beam) acts as a slow amplitude modulation on the interference signal. Consequently, the interference signal serves as an unbiased indicator of the path phase difference experienced by the write-out photon. As illustrated in Fig.~\ref{fig:phase detail}c, we experimentally verified the amplitude modulation using a fiber delay line. The fitted value of $\delta k$ was found to be $0.142$~cm$^{-1}$, allowing us to estimate $\delta\nu\approx678$~MHz, which closely matches our predetermined value. With this method, we can directly and unbiasedly lock the phase difference by employing a dual-band detuned locking beam. Crucially, the path difference drift only leads to a slow variation of the peak-to-peak value of the final interference signal, which hardly affects the phase locking point and can be easily compensated by a fiber delay line if necessary (see supplementary for details).

\begin{figure*}
	\centering
	\includegraphics[width=\textwidth]{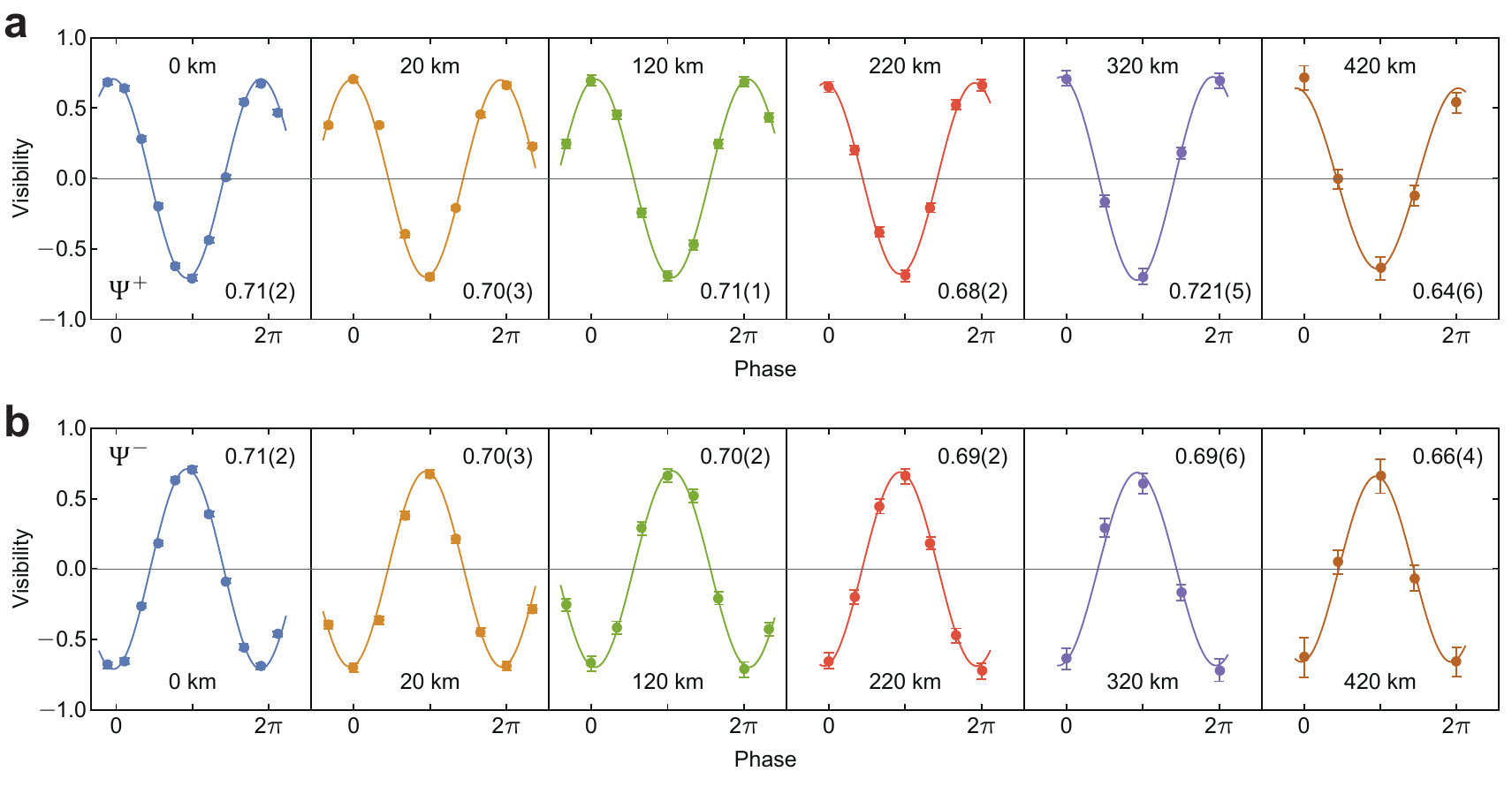}
	 \caption{\textbf{Verification of two-node entanglement.} 
	 The atomic modes are retrieved as optical modes and interfered at BS$_{\rm ro}$. The photon count visibility in two output modes of the BS oscillates as a function of the relative phase between the two read-out modes. The visibilities heralded by the write-out detector $D_1$ and $D_2$ are shown in (\textbf{a}) and (\textbf{b}) respectively, except the oscillation of 0~km which refers to the experiment that write-out photons are interfered and detected locally without QFC. The fitted visibilities are shown at the corner. Error bars indicate one standard deviation of the photon-counting statistics.
	 }
	\label{fig:visibility oscillation}
\end{figure*}
 
As shown in Fig.~\ref{fig:phase detail}d, we test the phase stabilization performance by introducing a weak pulse from BS$_{\rm ro}$, whose frequency is the same as the write-out photon. The pulses interfere at BS$_{\rm wo}$ and are detected by SNSPDs. Along with the scan of the locking phase difference, counts in two SNSPDs vary as a sinusoidal function, and we could deduce the visibility of different distances. At the test and atomic experiment of 0~km, interference appears at the local area without QFC, the locking method is similar to Fig.~\ref{fig:phase detail}a. Although the visibility decreases as the fiber length increases, it still remains above 90\% over 420~km. We also measure the phase residual errors, which are shown in Fig.~\ref{fig:phase detail}d. The measured visibilities are lower than the estimation from residual errors for several reasons. First, the interferometer requires re-locking time, and as the distance increases, the frequency of re-locking also rises, which reduces the measured visibility. This issue can be mitigated by extending the stroke of phase-locking devices. Additionally, imperfections in phase visibility measurements, such as intensity imbalance, also contribute to the discrepancy.

\section*{Experimental result}

After the challenges above are solved, we perform two-node entangling experiments for a series of fiber lengths. From the experiment of 120~km, we add a series of ultra-low loss coiled fiber to the write-out paths and the total fiber losses of two arms including delay line and deployed fibers are listed in Tab.~\ref{tab:concurrence}.

\begin{table*}[htbp]
	\centering
	\caption{\textbf{Summary of experimental results for different fiber lengths.}}
	\begin{tabular}[t]{C{2cm}C{2cm}C{2cm}C{2cm}C{2cm}C{2cm}C{2cm}C{2cm}}
	\toprule
	\multicolumn{2}{c}{Fiber length} &0~km&20~km&120~km&220~km&320~km&420~km\\
	\midrule
	\multicolumn{2}{c}{Fiber loss (dB)}&/&8.0&26.7&43.3&61.2&78.7\\
	\multicolumn{2}{c}{$\mathcal{C}(\Psi^+)$}&0.0470(56)&0.051(9)&0.048(3)&0.049(10)&0.052(7)&0.046(22)\\
	\multicolumn{2}{c}{SNR($\Psi^+$)}&32.2(4)&48.8(4)&48(1)&26(1)&15(2)&3.5(2)\\
	\multicolumn{2}{c}{$\mathcal{C}(\Psi^-)$}&0.0471(56)&0.046(9)&0.041(7)&0.050(11)&0.053(20)&0.016(22)\\
	\multicolumn{2}{c}{SNR($\Psi^-$)}&40.4(7)&46.7(6)&47(2)&44(4)&13(2)&2.6(2)\\
	\multicolumn{2}{c}{Entangling prob.}&$6.57\times10^{-2}$&$3.24\times10^{-3}$&$3.74\times10^{-4}$&$5.07\times10^{-5}$&$6.86\times10^{-6}$&$1.09\times10^{-6}$\\
	\bottomrule
	\end{tabular}
	\label{tab:concurrence}
\end{table*}

To evaluate the performance of our system, we read out the spin-waves back to photons 750~ns after the creation of atom-photon entanglement and measure the concurrence~\cite{chou_measurement-induced_2005} of two read-out modes in a delay choice fashion~\cite{ma_experimental_2012}. The concurrence can be expressed as $\mathcal{C}=\max(0,2|d|-2\sqrt{p_{00}p_{11}})$, where $p_{ij}$ is the probability of having i excitations in node Alice and j excitations in node Bob, $d=V(p_{01}+p_{10})/2$ and $V$ is the interference visibility of two modes. The probabilities $p_{ij}$ can be measured via photon statistics. To evaluate this interference visibility $V$, two read-out modes are combined via BS$_{\rm ro}$ and detected by two local single photon detectors. A relative phase is added between two read-out modes by the sandwich configuration of wave plates as shown in Fig.~\ref{fig:phase detail}a. The coincidence visibilities with two write-out modes are shown in Fig.~\ref{fig:visibility oscillation}. We fit the sinusoidal curves and deduce $V$ for concurrence calculation.

\begin{figure*}[htbp]
	\centering
	\includegraphics[width=0.65\textwidth]{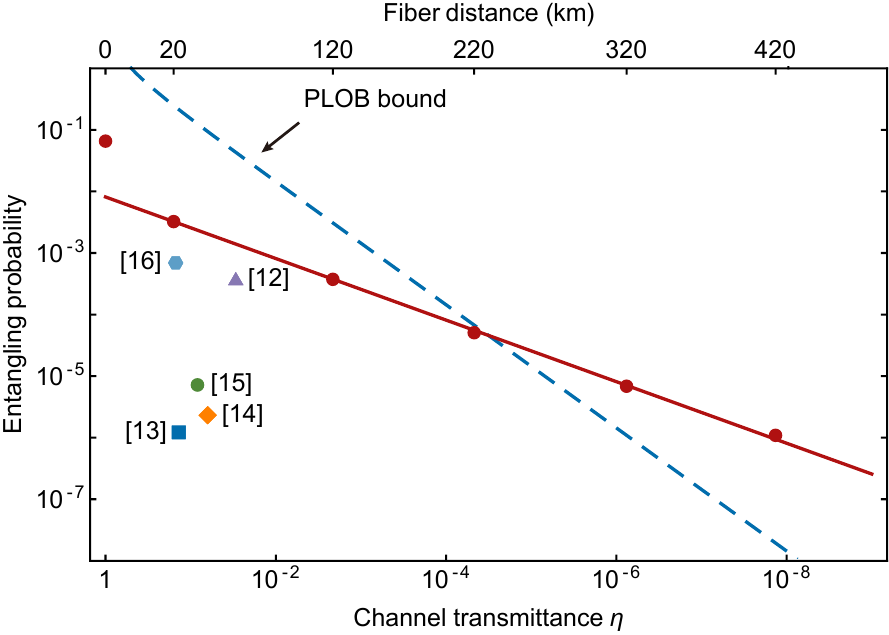}
	 \caption{\textbf{Entangling probabilities versus channel transmittance.} The entangling probabilities per trial are plotted against the channel transmittance. The red points refer to the experimental results at distances listed in Tab.~\ref{tab:concurrence}. The red solid line shows a linear fit of the result excluding the first point (0~km), showing a scaling of $\sqrt{\eta}$. Please note that the first point shows higher probability than the fit since it is measured in a different configuration without QFC. The blue dashed line shows the PLOB bound log$_2(1-\eta)$. The fiber distance is solely applicable to the current experiment.
 }
\label{fig:entanglement probability}
\end{figure*}

Tab.~\ref{tab:concurrence} also summarizes the concurrence results $\mathcal{C}$ of experiments with different fiber lengths. The entanglement can be verified even at 420~km, where we get a concurrence of $\mathcal{C}=0.046(22)$ heralded by the clicks of $D_1$ ($\Psi^+$). Besides these, we also get the entangling probability in one trial. As shown in the last row of Tab.~\ref{tab:concurrence} and Fig.~\ref{fig:entanglement probability}, we find that the memory-memory entangling probability scales with the square root of the channel transmittance ($\eta$). While for direction transmission of entangled photons, the entangling probability scales with $\eta$ instead, with an upper limit named as the Pirandola-Laurenza-Ottaviani-Banchi (PLOB) bound~\cite{pirandola_fundamental_2017}. As shown in Fig.~\ref{fig:entanglement probability}, when channel transmittance is lower than $10^{-4.5}$ (which refers to a distance over 230~km in our experiment), the entangling probability in our experiment is higher than the PLOB bound. Please note that in terms of quantum key distribution, the twin-field scheme~\cite{lucamarini_overcoming_2018} shows a similar scaling and beats the PLOB bound as well.

\section*{Discussion and outlook}

In summary, we have experimentally realized entanglement between two quantum memories via fibers up to 420~km. A more efficient quantum frequency conversion protocol helps us to reduce the fiber loss and extend the transmission distance. Additionally, we employ a novel phase stabilization scheme to ensure the implementation of the experiment based on the single-photon interference, which can be conveniently applied to other platforms. Furthermore, we demonstrate that the heralded entangling probability can beat the PLOB bound, which is an intrinsic constraint for direct entanglement distribution over lossy channel.

Through the integration of phase stabilization technology for separated nodes~\cite{liu_creation_2024}, our experiments can be implemented between nodes spatially separated by a similar distance. By leveraging Rydberg blockade mechanisms to generate atom-photon entanglement without higher-order events~\cite{li_entanglement_2013,sun_deterministic_2022}, the remote entanglement fidelity can be further improved significantly. Harnessing deterministic measurement of an ensemble qubit~\cite{yang_deterministic_2022}, this advance will enable more sophisticated applications like DI-QKD~\cite{zhang_device-independent_2022,nadlinger_experimental_2022}. Furthermore, our experiment also demonstrates the capability of high-rate entanglement generation at shorter distances ($\sim$100~km), which is of significant interest for the construction of quantum repeaters~\cite{briegel_repeater_1998,sangouard_quantum_2011}. Combining this feature with sub-second lifetime storage based on optical lattice~\cite{wang_cavity-enhanced_2021}, entanglement swapping can be employed to enable the connection of many such remote entanglement segments.

\section*{Acknowledgment}
\textbf{Funding}: This research was supported by the Innovation Program for Quantum Science and Technology (No.~2021ZD0301104, No.~2023ZD0300100), National Key R\&D Program of China (No.~2020YFA0309804), Anhui Initiative in Quantum Information Technologies, National Natural Science Foundation of China (No.~T2125010, No.~12274399), Chinese Academy of Sciences, USTC Research Funds of the Double First-Class Initiative (No.~YD9990002012), and the Postdoctoral Fellowship Program of CPSF (No.~GZB20240715). \textbf{Competing interests}: The authors declare no competing interests.

\setcounter{figure}{0}
\setcounter{table}{0}
\setcounter{equation}{0}

\onecolumngrid

\global\long\def\theequation{S\arabic{equation}}
\global\long\def\thefigure{S\arabic{figure}}
\global\long\def\thetable{S\arabic{table}}
\renewcommand{\arraystretch}{0.6}

\newpage

\newcommand{\msection}[1]{\vspace{\baselineskip}{\centering \textbf{#1}\\}\vspace{0.5\baselineskip}}

\msection{SUPPLEMENTAL MATERIAL}

\section{Details of the quantum frequency conversion}

\subsection{Structure of the QFC module}

We used the reverse-proton-exchange (RPE) periodically-poled lithium niobate (PPLN) waveguide chips to achieve the quantum frequency down-conversion from 780~nm to 1522~nm with a 1600~nm pump laser. An integrated waveguide structure as described in Ref.~\cite{yu2020} is used to couple the 1600~nm pump and 780~nm signal into the quasi-phase matching (QPM) waveguide. The QPM waveguide width is $\rm{8~\mu m}$ and the QPM period is $\rm{16.3~\mu m}$. The detailed waveguide structure is shown in Fig.~\ref{fig: QFC1}a. The total length of the RPE-PPLN waveguide chips is 56~mm with a mode filter and a tapered waveguide of 1~mm respectively. 

At the input port of the chip, the mode filter widths for the 780~nm signal and the 1600~nm pump are $\rm{2~\mu m}$ and $\rm{4~\mu m}$, respectively. The total transmittance, including fiber-to-waveguide coupling and propagation losses, is 79\% for the 780~nm signal and 65\% for the 1600~nm pump. A two-channel fiber array with HI780 fiber for 780~nm signal and SMF-28 Ultra fiber for 1600~nm pump was pigtailed to the input port. The center-to-center separation between two fiber channels is $\rm{127~\mu m}$. For the directional coupler, the waveguide width is $\rm{5.5~\mu m}$ and the edge-to-edge spacing is $\rm{3.5~\mu m}$. The measured coupling loss for the 780~nm signal light can be as low as 1\% and the coupling efficiency for the 1600~nm pump light can be as high as 98\%. At the output port of the chip, we also use a single-mode pigtail to couple the 1522~nm photons from the waveguide into a single-mode fiber with 90\% coupling efficiency. Following the integration of fiber pigtails on both input and output ports of the chip, the system achieves a measured external conversion efficiency of 64\%.

\subsection{Spectral filtering}
\begin{figure}[tbp]
	\centering
	\includegraphics[width=.7\linewidth]{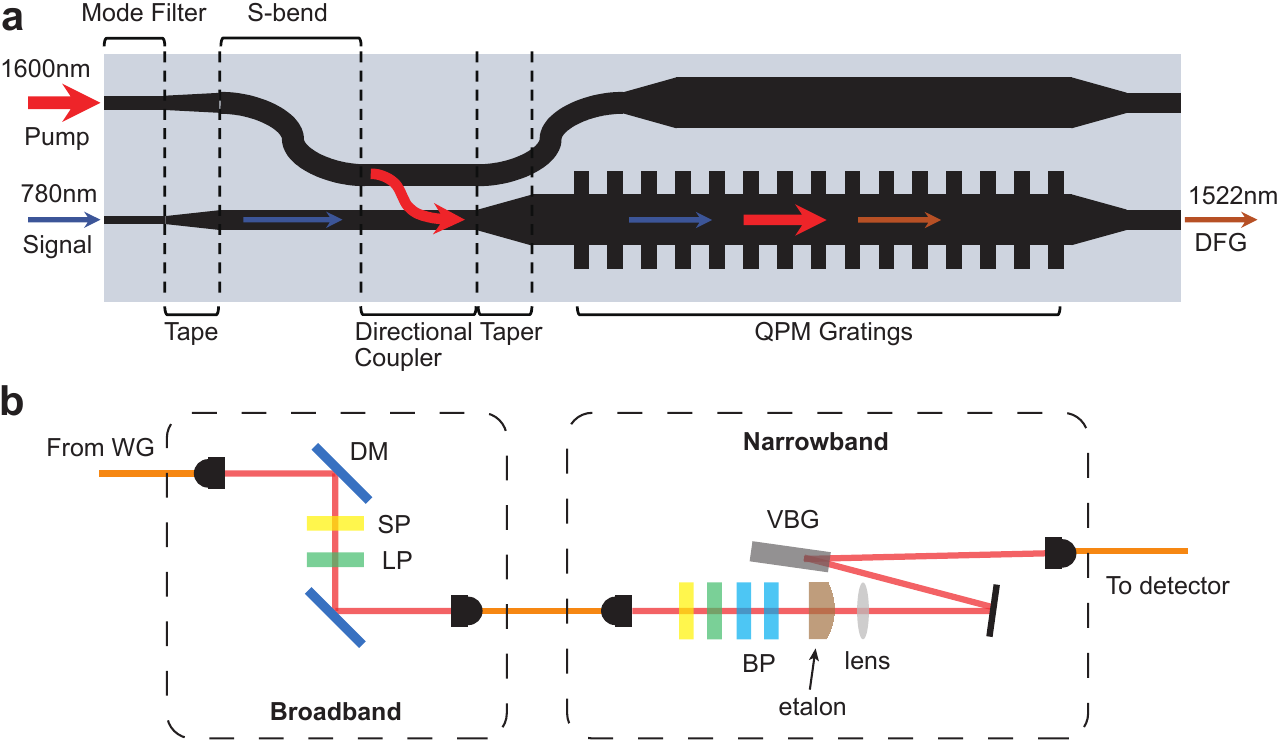}
	\caption{\textbf{a}. Detailed structure inside the PPLN waveguide chip. \textbf{b}. Detailed setup of two filter modules.}
	\label{fig: QFC1}
\end{figure}

\begin{figure}[tbp]
	\centering
	\includegraphics[width=.6\linewidth]{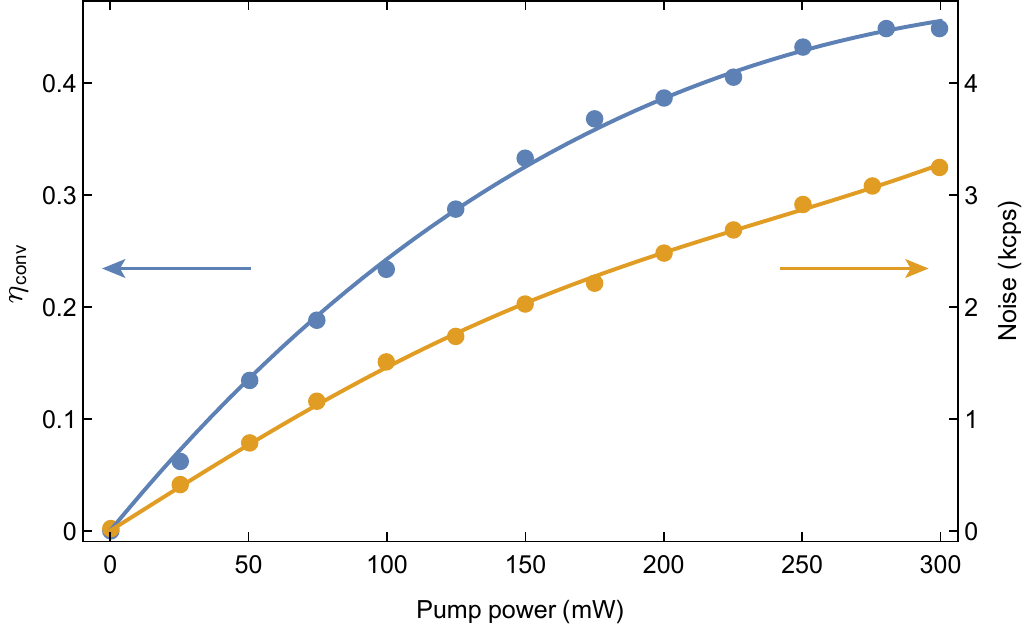}
	\caption{Performance of quantum frequency conversion.}
	\label{fig: QFC2}
\end{figure}

Amounts of noise are lead into signal during the QFC process, which limits the final signal-to-noise ratio (SNR). The broadband anti-Stokes Raman scattering (ASR) noise occupies the main position because of the small wavelength difference between the pump laser and the converted photon\cite{zaske2011,vanleent2020}. As discussed in the main text and shown in Fig.~\ref{fig: QFC1}b, the complete filter system consists of two main modules. The broadband filter module is placed right behind the waveguide. Two DMs (99.9\% reflectivity for 1522~nm and < 5\% reflectivity for 1600~nm)  and a short-pass filter (SP, 1570~nm cut-off, 99\% transmittance for 1522~nm) are used to remove the remnant pump. The short wave noise mainly caused by second and third harmonic generation is blocked by a long-pass filter (LP) edged at 1500~nm (99\% transmittance for 1522~nm). The residual ASR noise is removed by the narrow-band module, which consists of an etalon, a SP, a LP, a volume Bragg grating (VBG, a FWHM of 0.1~nm and a diffraction efficiency of 92\%) and two bandpass filters (BP, each has a FWHM of 1~nm and 97\% transmittance). The etalon has a FWHM of 70~MHz, a free-spectral range of 19~GHz and a transmission efficiency of 92\% at resonance. It is made from one piece of fused silica and we maintain the resonance point by stabilizing its temperature. A biconvex lens with f=150~mm is placed between the etalon and the VBG to match the spatial mode. We collect the signal twice after two modules with single mode fibers and the collection efficiencies are 88\% and 90\% respectively. The end-to-end efficiency (the collection efficiency of 88\% is excluded) and noise level of the QFC device with different pump laser power are shown in Fig.~\ref{fig: QFC2}. We set the power to 300~mW in the experiments and get an efficiency of 44\% with a noise level of 3.2~kHz. 

We measure and compare the correlation function $g^{(2)}(0)$ of write-out photons conditioned by detecting the corresponding read-out photon to verify the single-photon quality before and after the QFC process. The results are 0.377(3) and 0.36(5) respectively, which remain unchanged and prove that the quality can be well preserved.

\section{Details of phase stabilization}

\begin{figure}[htbp]
	\centering
	\includegraphics[width=.6\linewidth]{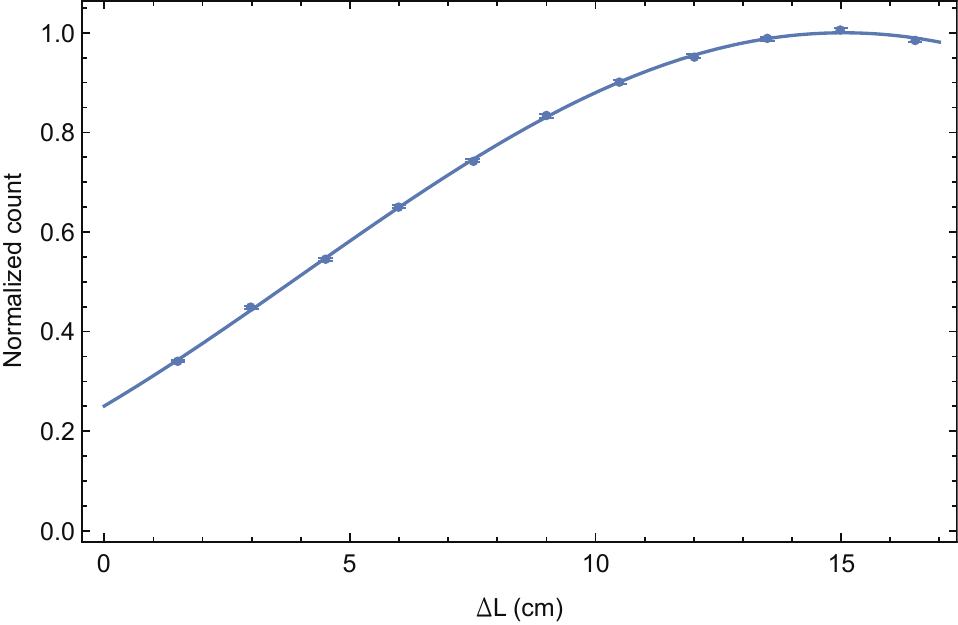}
	\caption{The probe photon count changes with $\Delta L$ by using a single detuned locking beam to stabilize the interferometer. 
	}
	\label{fig: delayline}
\end{figure}

We review the process of the entanglement creation and verification. A write pulse induces a collective excitation and a write-out scattered photon which form a Fock state entanglement. Then the atomic state is retrieved as a read-out photon for verification. The local entanglement between write-out and read-out photon mode reads
\begin{equation}
	\ket{00}+\sqrt{\chi}e^{i\varphi_{wr}}\ket{1_{\rm wo}1_{\rm ro}},
\end{equation}
where
\begin{equation}
	\varphi_{wr}=\phi_w+\phi_{\rm wo}+\phi_r+\phi_{\rm ro}.
\end{equation}
$\phi_i$ refers to the path phase of the write pulse (w), write-out photon (wo), read pulse (r) and read-out photon (ro) process. Here we neglect the initial phase of lasers because the same resources are used in two nodes. When a click event happens at one of the SNSPDs, the entanglement of read-out modes is
\begin{equation}
	\ket{\Psi_{\rm ro}^{\pm}}=(e^{i\varphi_{wr}^A}\ket{10}_{AB}\pm e^{i\varphi_{wr}^B}\ket{01}_{AB})/\sqrt{2}.
\end{equation}
During the entanglement verification, we combine two read-out modes at $BS_{\rm ro}$ to measure visibility V. We should ensure the phase $\Delta\varphi_{wr} (=\varphi_{wr}^A-\varphi_{wr}^B)$ is stabilized. We divide the total phase difference $\Delta\varphi_{wr}$ into two parts
\begin{equation}
	\Delta\varphi_{wr}=\Delta(\phi_w+\phi_r)+\Delta(\phi_{\rm wo}+\phi_{\rm ro}),
\end{equation}
and configure two interferometers to stabilize them respectively while measuring visibility V in the main text. Here we focus on the method to stabilize the write-out-read-out interferometer. $\Delta(\phi_{\rm wo}+\phi_{\rm ro})$ are given by:
\begin{equation}
	\Delta(\phi_{\rm wo}+\phi_{\rm ro})=\Delta(k_{\rm ro}L_{\rm ro})+\Delta(k_{\rm wo}L_{\rm wo})-\Delta(k_{p}L_{p})+\Delta(k_{tel}L_{tel}),
\end{equation}
where $k_i$ and $L_i$ are the wave number and the transmission distance of the read-out (ro) photon, write-out (wo) photon, pump (p) laser and telecom (tel) field after conversion. We have the energy conservation principle in QFC process $k_{tel}=k_{\rm wo}-k_{p}$, so
\begin{equation}\label{eq: real_phase}
	\Delta(\phi_{\rm wo}+\phi_{\rm ro})=\Delta(k_{\rm ro}L_{\rm ro})+\Delta(k_{\rm wo}L_{wol})-\Delta(k_{p}L_{pl}),
\end{equation}
where $L_{wol}=L_{\rm wo}+L_{tel}$, $L_{pl}=L_{p}+L_{tel}$. In the former experiment\cite{yu2020}, we introduce a locking beam ($k_{lo}$), which is far detuned from the write-out photon, from $BS_{\rm ro}$ to stabilize the interferometer. It gives
\begin{equation}\label{eq: lock_phase_1}
	\Delta(k_{lo}L_{\rm ro})+\Delta(k_{lo}L_{wol})-\Delta(k_{p}L_{pl})=constant.
\end{equation}
We compare it with Eq.~\ref{eq: real_phase}, and find that the residual uncertain phase is
\begin{equation}
	\varphi_{rs}=(k_{\rm ro}-k_{lo})\Delta L_{\rm ro}+(k_{\rm wo}-k_{lo})\Delta L_{wol}.
\end{equation}
This indicates that the slow drift of this uncertain phase comes from two aspects: wave number differences and distance difference between the channels. The former mainly originates from laser sources. We lock all lasers onto a stable and narrow-linewidth reference cavity to eliminate their drift. In our experiments, $k_{\rm ro}-k_{lo}\approx215~\mathrm{m}^{-1}$, $k_{\rm wo}-k_{lo}\approx21.2~\mathrm{m}^{-1}$, $L_{\rm ro}\approx 10~\mathrm{m}$, and the thermal expansion coefficient of the fiber is about $5.5\times10^{-7}~\mathrm{K}^{-1}$. If we use the drift of $L_{\rm ro}$ to evaluate the drift of $\Delta(L_{\rm ro})$, we get: $215~\mathrm{m}^{-1}\times10~\mathrm{m}\times5.5\times10^{-7}~\mathrm{K}^{-1}\approx0.0012~\mathrm{Rad/K}$, which
means the slow drift of read-out fiber can be neglected at room temperature.
As for the write-out fiber: $L_{\rm wo}\approx 100~\mathrm{km}$, we get $21.2~\mathrm{m}^{-1}\times100~\mathrm{km}\times5.5\times10^{-7}~\mathrm{K}^{-1}\approx1.16~\mathrm{Rad/K}$, which means the drift can lead to a considerable effect on the locking point even if a slight temperature difference of 0.2~K appears between two fiber channels. As shown in Fig.~\ref{fig: delayline}, we change $\Delta L$ actively by using a delay line and use the same method as Fig.~2d in the main text to observe the change of locking phase. Here we prove that this problem can be solved by using a locking beam with two frequency components. 

To simplify the following derivation, we only focus on the difference between the write-out channels. The notations and their corresponding descriptions are listed in Tab.~\ref{tab: symbols}.
\begin{table*}[ht]
	\centering
	\caption{A summary of notations and their descriptions in the derivation. For the distance and amplitude notations, we use the superscript 1, 2 for referring to the channels from Alice and Bob respectively.}
	\begin{tabular}[t]{C{2cm} C{10cm}}
		\toprule
		Notations & Description \\
		\midrule
		$\nu_{+(-)}$& The frequency of the blue (red) detuning beam.\\
		$\nu_{0}$& The frequency of the write-out photon.\\
		$\Delta\nu$& $\nu_+-\nu_-$\\
		$\delta\nu$& $\nu_+-\nu_0$ or $\nu_0-\nu_-$\\
		\\
		$k_{+(-)}$& The wave number of the blue (red) detuning beam.\\
		$k_{0}$& The wave number of the write-out photon.\\
		$\Delta k$& $k_+-k_-$\\
		$\delta k$& $k_+-k_0$ or $k_0-k_-$\\
		\\
		$\omega_{+(-)}$& The angular frequency of the blue (red) detuning beam.\\
		$\omega_{0}$& The angular frequency of the write-out photon.\\
		$\Delta\omega$& $\omega_+-\omega_-$\\
		\\
		$\psi_{+(-)}$& The original phase of the blue (red) detuning beam.\\
		\\
		$A(B)$& The amplitude of the blue (red) detuning beam\\
		$L$& The length of the write-out fiber\\
		$\Delta L$& $L_1-L_2$\\
		$n_0$& The refractive index of the write-out photon in the fiber\\
		$\delta n$& The refractive index difference between the blue detuning beam and the write-out photon\\
		\bottomrule
	\end{tabular}
	\label{tab: symbols}
\end{table*}
First, we combine two detuned beams with a beam splitter and inject them into the write-out-read-out interferometer. The total amplitude behind the $BS_{\rm wo}$ reads
\begin{equation}
	A_1e^{-i(\omega_+t-k_+L_1-\psi_+)}+A_2e^{-i(\omega_+t-k_+L_2-\psi_+)}+B_1e^{-i(\omega_-t-k_-L_1-\psi_-)}+B_2e^{-i(\omega_-t-k_-L_2-\psi_-)}.
\end{equation}
Then we can calculate and expand the detected intensity at the APD,
\begin{equation}
	\begin{split}
		I=&(A_1+A_2)^2\cos^2\frac{1}{2}k_+\Delta L+(A_1-A_2)^2\sin^2\frac{1}{2}k_+\Delta L\\
		+&(B_1+B_2)^2\cos^2\frac{1}{2}k_-\Delta L+(B_1-B_2)^2\sin^2\frac{1}{2}k_-\Delta L\\
		+&2(A_1+A_2)(B_1+B_2)\cos(-\Delta\omega t+\frac{1}{2}\Delta k(L_1+L_2)+\Delta\psi)\cos\frac{1}{2}k_+\Delta L\cos\frac{1}{2}k_-\Delta L\\
		+&2(A_1-A_2)(B_1-B_2)\cos(-\Delta\omega t+\frac{1}{2}\Delta k(L_1+L_2)+\Delta\psi)\sin\frac{1}{2}k_+\Delta L\sin\frac{1}{2}k_-\Delta L\\
		+&2(A_1+A_2)(B_1-B_2)\sin(-\Delta\omega t+\frac{1}{2}\Delta k(L_1+L_2)+\Delta\psi)\cos\frac{1}{2}k_+\Delta L\sin\frac{1}{2}k_-\Delta L\\
		+&2(A_1-A_2)(B_1+B_2)\sin(-\Delta\omega t+\frac{1}{2}\Delta k(L_1+L_2)+\Delta\psi)\sin\frac{1}{2}k_+\Delta L\cos\frac{1}{2}k_-\Delta L,\\
	\end{split}
\end{equation}
where $\Delta\omega=\omega_+-\omega_-\approx1.3~\mathrm{GHz}$, so the last four terms oscillate with high frequency and have no effect on the APD with a bandwidth of megahertz. We neglect them and simplify the first four terms,
\begin{equation}\label{eq: intensity_1}
	I=S+2(A_1A_2+B_1B_2)\cos\overline{k}\Delta L\cos\frac{1}{2}\Delta k\Delta L+2(A_1A_2-B_1B_2)\sin\overline{k}\Delta L\sin\frac{1}{2}\Delta k\Delta L,
\end{equation}
where $S=A_1^2+A_2^2+B_1^2+B_2^2=constant$ and $\overline{k}=(k_++k_-)/2$. In our experiments, two detuned beams are set to the same intensity, which means $A_1A_2-B_1B_2\approx 0$. Here we only consider the second term and assume that the refractive index $n$ changes linearly with the laser frequency, which implies
\begin{equation}
	k_{\pm}=\frac{2\pi n_{\pm}\nu_{\pm}}{c}=\frac{2\pi}{c}(n_0\pm\delta n)(\nu_0\pm\delta\nu),
\end{equation}
where $c$ is the speed of light in the vacuum. Then we can get
\begin{equation}
	\begin{split}
		\overline{k}&=\frac{2\pi}{c}(n_0\nu_0+\delta n\delta\nu)=k_0+\frac{2\pi}{c}\delta n\delta\nu\approx k_0,\\
		\frac{1}{2}\Delta k&=\frac{2\pi}{c}(n_0\delta\nu+\nu_0\delta n)=\delta k\ll k_0.
	\end{split}
\end{equation}
We substitute them into the Eq.~\ref{eq: intensity_1} and it gives
\begin{equation}\label{eq: intensity_2}
	I\approx S+\tilde{A}\cos k_0\Delta L,
\end{equation}
where $\tilde{A}=2(A_1A_2+B_1B_2)\cos\delta k\Delta L$, it can be treated as a slow amplitude modulation on the interference signal and hardly affects on the locking phase. So the interference signal $I$ can be used to characterize the phase difference of the write-out photon. As shown in Fig.~2c in the main text, we observe the sinusoidal variation of the peak-to-peak value of the interference signal as $\Delta L$ changes. The fitted $\delta k$ is $0.142~\mathrm{cm}^{-1}$, which yields an estimated frequency detuning $\delta\nu\approx678~\mathrm{MHz}$ (the delay line uses air as medium and the parameters we use here: $n\approx1$, $\delta n\approx 0$), demonstrating excellent agreement with our predetermined parameter. We record the peak-to-peak value and compensate it by using the fiber delay line every 15 or 30 minutes.
 
\section{Experimental imperfection of V}
\subsection{Decrease of SNR}

As listed in Tab.~1, as the fiber loss increases in long-distance scenarios, dark counts from SNSPDs emerge as the dominant noise source. This occurs because the QFC noise experience identical attenuation to the signal photons during propagation.

When we measure the interference visibility of the read-out field, the click events heralded by random write-out noise contribute equally to two read-out interference outputs, which gives
\begin{equation}
	\begin{split}
	V&=\frac{p_{max}+p_{noise}/2-p_{min}-p_{noise}/2}{p_{max}+p_{noise}/2+p_{min}+p_{noise}/2}=\frac{p_{max}-p_{min}}{p_{max}+p_{min}+p_{noise}}\\
	&=V_0\frac{co_{\rm wo}}{co_{\rm wo}+co_{no}}=V_0V_{\rm SNR},
	\end{split}
\end{equation}
where $co_{wo(no)}$ refers to the coincidence probability with the real write-out (noise) event. We can estimate $co_{\rm wo}\approx p_{\rm wo}\eta_{ret}$ and $co_{no}\approx p_{no}p_{\rm ro}$, where $p_{\rm wo}$ is the probability of a real write-out event at one SNSPD, $p_{no}=p_{\rm wo}/SNR$ is the noise in the SNSPD, $p_{\rm ro}$ is the probability of read-out photons. After applying the parameters in our experiments, the decreases that V suffers for different distances are listed in Tab.~\ref{tab: SNR}.

\begin{table*}[ht]
	\centering
	\caption{A summary of SNR and $V_{\rm SNR}$.}
	\begin{tabular}[t]{C{1cm} C{1cm} C{1.5cm} C{1.5cm} C{1.5cm} C{1.5cm} C{1.5cm} C{1.5cm}}
		\toprule
		\multicolumn{2}{c}{Distance (km)}& 0&20&120&220&320&420 \\
		\midrule
		\multirow{2}*{$\Psi^+$}&SNR&32.2&48.8&48&26&15&3.5\\
		&$V_{\rm SNR}$&0.993&0.995&0.995&0.991&0.984&0.936\\
		\midrule
		\multirow{2}*{$\Psi^-$}&SNR&40.4&46.7&47&44&13&2.6\\
		&$V_{\rm SNR}$&0.994&0.995&0.995&0.995&0.982&0.915\\
		\bottomrule
	\end{tabular}
	\label{tab: SNR}
\end{table*}

\subsection{Mismatch of write-out field}
The identity of write-out fields is the key requirement for the entanglement-building process. A significant cause of decreasing the identity is the difference in arrival time. We align the fields from two memory nodes carefully to reduce its effect. To evaluate the difference, we accumulate and mark the arrival time of each count from two optical paths respectively. Here the mismatches for different distances and the value to describe the decrease of V are listed in Tab.~\ref{tab: mismatch}. 

\begin{table*}[ht]
	\centering
	\caption{A summary of mismatch and $V_{M}$.}
	\begin{tabular}[t]{C{2.5cm} C{1.5cm} C{1.5cm} C{1.5cm} C{1.5cm} C{1.5cm} C{1.5cm}}
		\toprule
		Distance (km)& 0&20&120&220&320&420 \\
		\midrule
		Mismatch (ns)&2.22&1.47&2.07&1.81&0.50&7.56\\
		$V_{M}$&0.9980&0.9992&0.9984&0.9987&0.9999&0.9832\\
		\bottomrule
	\end{tabular}
	\label{tab: mismatch}
\end{table*}

\subsection{Photon indistinguishability}
\begin{figure}[htbp]
	\centering
	\includegraphics[width=.4\linewidth]{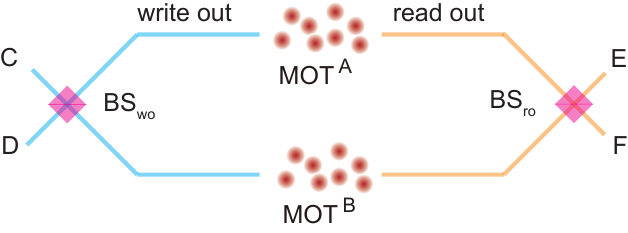}
	\caption{Interference of the write-out and read-out fields.}
	\label{fig: indistinguishability}
\end{figure}

We first consider the interference of the write-out field. As shown in Fig.~\ref{fig: indistinguishability}, two entanglement states from Alice and Bob can be expressed as
\begin{equation}
	\ket{\Psi_A}\ket{\Psi_B}=\ket{1_A0_B}_a\ket{1_A0_B}_{\rm wo}\ket{\psi}_{\rm wo}+e^{i\theta}\ket{0_A1_B}_a\ket{0_A1_B}_{\rm wo}(\sqrt{\eta_{\rm wo}}\ket{\psi}_{\rm wo}+\sqrt{1-\eta_{\rm wo}}\ket{\psi^\perp}_{\rm wo}),
\end{equation}
where the subscript $a$ means the atomic state, while $wo$ means the write-out photon state. $\ket{\psi}$ is the other degree of freedom of photons, and $\braket{\psi|\psi^\perp}=0$. $\eta_{\rm wo(ro)}$ represents the indistinguishability parameter of the write-out (read-out) field. Then we consider the event at output C:
\begin{equation}
	\ket{\Psi_C}=(\ket{1_A0_B}_a\ket{\psi}_{\rm wo}+e^{i\theta}\ket{0_A1_B}_a(\sqrt{\eta_{\rm wo}}\ket{\psi}_{\rm wo}+\sqrt{1-\eta_{\rm wo}}\ket{\psi^\perp}_{\rm wo}))\ket{1_C0_D}_{\rm wo}.
\end{equation}
The states of atoms are then retrieved as photons and interfered at $BS_{\rm ro}$, it gives
\begin{equation}
	\begin{split}
		\ket{\Psi_{\rm CE}}&=\ket{1_C0_D}\ket{1_E0_F}(\ket{\psi}_{\rm wo}\ket{\psi}_{\rm ro}+e^{i\theta}(\sqrt{\eta_{\rm wo}}\ket{\psi}_{\rm wo}+\sqrt{1-\eta_{\rm wo}}\ket{\psi^{\perp}}_{\rm wo})(\sqrt{\eta_{\rm ro}}\ket{\psi}_{\rm ro}+\sqrt{1-\eta_{\rm ro}}\ket{\psi^{\perp}}_{\rm ro}))\\
		\ket{\Psi_{\rm CF}}&=\ket{1_C0_D}\ket{0_E1_F}(\ket{\psi}_{\rm wo}\ket{\psi}_{\rm ro}-e^{i\theta}(\sqrt{\eta_{\rm wo}}\ket{\psi}_{\rm wo}+\sqrt{1-\eta_{\rm wo}}\ket{\psi^{\perp}}_{\rm wo})(\sqrt{\eta_{\rm ro}}\ket{\psi}_{\rm ro}+\sqrt{1-\eta_{\rm ro}}\ket{\psi^{\perp}}_{\rm ro})).
	\end{split}
\end{equation}
So the visibility can be given by $\sqrt{\eta_{\rm wo}\eta_{\rm ro}}$. Here we use Hong-Ou-Mandel (HOM) interference to estimate the indistinguishability between write-out fields from two nodes as well as two read-out fields\cite{tsujimoto2021,liu2024}
\begin{equation}
	g_{\rm HOM}^{(2)}=\frac{g_A^{(2)}+\zeta^2g_B^{(2)}+2(1-\eta)\zeta}{(1+\zeta)^2},
\end{equation}
where $\zeta$ is the relative intensity of two fields, and we set $\zeta=1$. The measured $g^{(2)}$, $\eta_{wo(ro)}$ and the value $V_I$ that describes the decrease of V are listed at Tab.~\ref{tab: indistinguishability}. $g^{(2)}_{\rm wo(ro)}$ are measured conditioned by detecting the corresponding read-out (write out) photon.

\begin{table*}[ht]
	\centering
	\caption{$g^{(2)}$, $\eta$ and $V_{I}$.}
	\begin{tabular}[t]{C{1.5cm}C{1.5cm} C{1.5cm} C{2cm} C{1.5cm} C{1.5cm} C{1.5cm}C{2cm}C{2cm}}
		\toprule
		$g^{(2)}_{\rm wo,A}$&$g^{(2)}_{\rm wo,B}$&$g^{(2)}_{\rm wo,HOM}$&$\eta_{\rm wo}$&$g^{(2)}_{\rm ro,A}$&$g^{(2)}_{\rm ro,B}$&$g^{(2)}_{\rm ro,HOM}$&$\eta_{\rm ro}$&$V_I$ \\
		\midrule
		0.387(11)&0.348(11)&0.210(28)&0.95(6)&0.302(10)&0.397(14)&0.215(28)&0.92(6)&0.93(4)\\
		\bottomrule
	\end{tabular}
	\label{tab: indistinguishability}
\end{table*}

\subsection{High order excitations}
In the measurement of visibility, the read-out photon interference also contains the contribution of high order excitation. The accidental click events caused by these uncorrelated excitations can also give equal contributions to both outputs. We use the equation S35 in the supplementary of Ref.~\cite{liu2024} to evaluate the max visibility $V_H=1/(1+2\chi(3-2\eta_r))=0.769$, where internal excitation probability $\chi\approx0.06$ and the mean retrieval efficiency $\eta_r\approx0.25$. 

\subsection{Summary}
Finally, we summarize all experimental imperfections discussed above and remote phase locking visibility $V_{P}$ in the main text in Tab.~\ref{tab: result}. After multiplying all of them, we can get the theory value $V_{\rm theory}$ to compare with the experimental results for different distances. All experiment results are close to the theory value. Some extra tests of imperfection are conducted after all main entanglement experiments have been done. The state of the whole system is not at its best at this time, which leads to a slightly lower result of the theory evaluation. The main limitation of the visibility comes from the high-order excitation, which can be solved by using the Rydberg blockade mechanism or other methods. Nonetheless, we can easily lower the excitation rate to decrease its effect. For example, at the experiment of 320~km and 420~km, if we lower the rate to 2\%, the final visibility can promote to 0.78(5) and 0.72(5) though the SNR and probability of entanglement creation fall to a third.
\begin{table*}[ht]
	\centering
	\caption{The comparison between the theoretical estimation of V and the experimental results.}
	\begin{tabular}[t]{C{1cm}C{1cm} C{1.5cm} C{1.5cm} C{1.5cm} C{1.5cm} C{1.5cm} C{1.5cm}}
		\toprule
		\multicolumn{2}{c}{Distance (km)}& 0&20&120&220&320&420 \\
		\midrule
		\multirow{7}*{$\Psi^+$}&$V_{\rm SNR}$&0.993&0.995&0.995&0.991&0.984&0.936\\
		&$V_{M}$&0.9980&0.9992&0.9984&0.9987&0.9999&0.9832\\
		&$V_{I}$&0.93(4)&0.93(4)&0.93(4)&0.93(4)&0.93(4)&0.93(4)\\
		&$V_{H}$&0.769&0.769&0.769&0.769&0.769&0.769\\
		&$V_{P}$&0.98(1)&0.96(1)&0.96(2)&0.95(1)&0.94(2)&0.93(4)\\
		\cmidrule(r){2-8}
		&$V_{\rm theory}$&0.70(3)&0.68(3)&0.68(3)&0.67(3)&0.66(3)&0.61(4)\\
		&$V_{\rm exp}$&0.71(2)&0.70(3)&0.71(1)&0.68(2)&0.721(5)&0.64(6)\\
		\midrule
		\multirow{7}*{$\Psi^-$}&$V_{\rm SNR}$&0.994&0.995&0.995&0.995&0.982&0.915\\
		&$V_{M}$&0.9980&0.9992&0.9984&0.9987&0.9999&0.9832\\
		&$V_{I}$&0.93(4)&0.93(4)&0.93(4)&0.93(4)&0.93(4)&0.93(4)\\
		&$V_{H}$&0.769&0.769&0.769&0.769&0.769&0.769\\
		&$V_{P}$&0.98(1)&0.96(1)&0.96(2)&0.95(1)&0.94(2)&0.93(4)\\
		\cmidrule(r){2-8}
		&$V_{\rm theory}$&0.70(3)&0.68(3)&0.68(3)&0.68(3)&0.66(3)&0.60(4)\\
		&$V_{\rm exp}$&0.71(2)&0.70(3)&0.70(2)&0.69(2)&0.69(6)&0.66(4)\\
		\bottomrule
	\end{tabular}
	\label{tab: result}
\end{table*}


\begin{thebibliography}{0}%
\makeatletter
\providecommand \@ifxundefined [1]{%
 \@ifx{#1\undefined}
}%
\providecommand \@ifnum [1]{%
 \ifnum #1\expandafter \@firstoftwo
 \else \expandafter \@secondoftwo
 \fi
}%
\providecommand \@ifx [1]{%
 \ifx #1\expandafter \@firstoftwo
 \else \expandafter \@secondoftwo
 \fi
}%
\providecommand \natexlab [1]{#1}%
\providecommand \enquote  [1]{``#1''}%
\providecommand \bibnamefont  [1]{#1}%
\providecommand \bibfnamefont [1]{#1}%
\providecommand \citenamefont [1]{#1}%
\providecommand \href@noop [0]{\@secondoftwo}%
\providecommand \href [0]{\begingroup \@sanitize@url \@href}%
\providecommand \@href[1]{\@@startlink{#1}\@@href}%
\providecommand \@@href[1]{\endgroup#1\@@endlink}%
\providecommand \@sanitize@url [0]{\catcode `\\12\catcode `\$12\catcode
  `\&12\catcode `\#12\catcode `\^12\catcode `\_12\catcode `\%12\relax}%
\providecommand \@@startlink[1]{}%
\providecommand \@@endlink[0]{}%
\providecommand \url  [0]{\begingroup\@sanitize@url \@url }%
\providecommand \@url [1]{\endgroup\@href {#1}{\urlprefix }}%
\providecommand \urlprefix  [0]{URL }%
\providecommand \Eprint [0]{\href }%
\providecommand \doibase [0]{https://doi.org/}%
\providecommand \selectlanguage [0]{\@gobble}%
\providecommand \bibinfo  [0]{\@secondoftwo}%
\providecommand \bibfield  [0]{\@secondoftwo}%
\providecommand \translation [1]{[#1]}%
\providecommand \BibitemOpen [0]{}%
\providecommand \bibitemStop [0]{}%
\providecommand \bibitemNoStop [0]{.\EOS\space}%
\providecommand \EOS [0]{\spacefactor3000\relax}%
\providecommand \BibitemShut  [1]{\csname bibitem#1\endcsname}%
\let\auto@bib@innerbib\@empty
\end{thebibliography}%


\begin{thebibliography}{50}%
	\makeatletter
	\providecommand \@ifxundefined [1]{%
	 \@ifx{#1\undefined}
	}%
	\providecommand \@ifnum [1]{%
	 \ifnum #1\expandafter \@firstoftwo
	 \else \expandafter \@secondoftwo
	 \fi
	}%
	\providecommand \@ifx [1]{%
	 \ifx #1\expandafter \@firstoftwo
	 \else \expandafter \@secondoftwo
	 \fi
	}%
	\providecommand \natexlab [1]{#1}%
	\providecommand \enquote  [1]{``#1''}%
	\providecommand \bibnamefont  [1]{#1}%
	\providecommand \bibfnamefont [1]{#1}%
	\providecommand \citenamefont [1]{#1}%
	\providecommand \href@noop [0]{\@secondoftwo}%
	\providecommand \href [0]{\begingroup \@sanitize@url \@href}%
	\providecommand \@href[1]{\@@startlink{#1}\@@href}%
	\providecommand \@@href[1]{\endgroup#1\@@endlink}%
	\providecommand \@sanitize@url [0]{\catcode `\\12\catcode `\$12\catcode `\&12\catcode `\#12\catcode `\^12\catcode `\_12\catcode `\%12\relax}%
	\providecommand \@@startlink[1]{}%
	\providecommand \@@endlink[0]{}%
	\providecommand \url  [0]{\begingroup\@sanitize@url \@url }%
	\providecommand \@url [1]{\endgroup\@href {#1}{\urlprefix }}%
	\providecommand \urlprefix  [0]{URL }%
	\providecommand \Eprint [0]{\href }%
	\providecommand \doibase [0]{https://doi.org/}%
	\providecommand \selectlanguage [0]{\@gobble}%
	\providecommand \bibinfo  [0]{\@secondoftwo}%
	\providecommand \bibfield  [0]{\@secondoftwo}%
	\providecommand \translation [1]{[#1]}%
	\providecommand \BibitemOpen [0]{}%
	\providecommand \bibitemStop [0]{}%
	\providecommand \bibitemNoStop [0]{.\EOS\space}%
	\providecommand \EOS [0]{\spacefactor3000\relax}%
	\providecommand \BibitemShut  [1]{\csname bibitem#1\endcsname}%
	\let\auto@bib@innerbib\@empty
	\bibitem [{\citenamefont {Kimble}(2008)}]{kimble_internet_2008}%
	  \BibitemOpen
	  \bibfield  {author} {\bibinfo {author} {\bibfnamefont {H.~J.}\ \bibnamefont {Kimble}},\ }\bibfield  {title} {\bibinfo {title} {The quantum internet},\ }\href {https://doi.org/10.1038/nature07127} {\bibfield  {journal} {\bibinfo  {journal} {Nature}\ }\textbf {\bibinfo {volume} {453}},\ \bibinfo {pages} {1023} (\bibinfo {year} {2008})}\BibitemShut {NoStop}%
	\bibitem [{\citenamefont {Wehner}\ \emph {et~al.}(2018)\citenamefont {Wehner}, \citenamefont {Elkouss},\ and\ \citenamefont {Hanson}}]{wehner_quantum_2018}%
	  \BibitemOpen
	  \bibfield  {author} {\bibinfo {author} {\bibfnamefont {S.}~\bibnamefont {Wehner}}, \bibinfo {author} {\bibfnamefont {D.}~\bibnamefont {Elkouss}},\ and\ \bibinfo {author} {\bibfnamefont {R.}~\bibnamefont {Hanson}},\ }\bibfield  {title} {\bibinfo {title} {Quantum internet: {A} vision for the road ahead},\ }\href {https://doi.org/10.1126/science.aam9288} {\bibfield  {journal} {\bibinfo  {journal} {Science}\ }\textbf {\bibinfo {volume} {362}},\ \bibinfo {pages} {eaam9288} (\bibinfo {year} {2018})}\BibitemShut {NoStop}%
	\bibitem [{\citenamefont {Jiang}\ \emph {et~al.}(2007)\citenamefont {Jiang}, \citenamefont {Taylor}, \citenamefont {S\o{}rensen},\ and\ \citenamefont {Lukin}}]{jiang_distributedcomputation_2007}%
	  \BibitemOpen
	  \bibfield  {author} {\bibinfo {author} {\bibfnamefont {L.}~\bibnamefont {Jiang}}, \bibinfo {author} {\bibfnamefont {J.~M.}\ \bibnamefont {Taylor}}, \bibinfo {author} {\bibfnamefont {A.~S.}\ \bibnamefont {S\o{}rensen}},\ and\ \bibinfo {author} {\bibfnamefont {M.~D.}\ \bibnamefont {Lukin}},\ }\bibfield  {title} {\bibinfo {title} {Distributed quantum computation based on small quantum registers},\ }\href {https://doi.org/10.1103/PhysRevA.76.062323} {\bibfield  {journal} {\bibinfo  {journal} {Physical Review A}\ }\textbf {\bibinfo {volume} {76}},\ \bibinfo {pages} {062323} (\bibinfo {year} {2007})}\BibitemShut {NoStop}%
	\bibitem [{\citenamefont {Gottesman}\ \emph {et~al.}(2012)\citenamefont {Gottesman}, \citenamefont {Jennewein},\ and\ \citenamefont {Croke}}]{gottesman_sensing_2012}%
	  \BibitemOpen
	  \bibfield  {author} {\bibinfo {author} {\bibfnamefont {D.}~\bibnamefont {Gottesman}}, \bibinfo {author} {\bibfnamefont {T.}~\bibnamefont {Jennewein}},\ and\ \bibinfo {author} {\bibfnamefont {S.}~\bibnamefont {Croke}},\ }\bibfield  {title} {\bibinfo {title} {Longer-baseline telescopes using quantum repeaters},\ }\href {https://doi.org/10.1103/PhysRevLett.109.070503} {\bibfield  {journal} {\bibinfo  {journal} {Physical Review Letters}\ }\textbf {\bibinfo {volume} {109}},\ \bibinfo {pages} {070503} (\bibinfo {year} {2012})}\BibitemShut {NoStop}%
	\bibitem [{\citenamefont {K{\'o}m{\'a}r}\ \emph {et~al.}(2014)\citenamefont {K{\'o}m{\'a}r}, \citenamefont {Kessler}, \citenamefont {Bishof}, \citenamefont {Jiang}, \citenamefont {S{\o}rensen}, \citenamefont {Ye},\ and\ \citenamefont {Lukin}}]{komar_clock_2014}%
	  \BibitemOpen
	  \bibfield  {author} {\bibinfo {author} {\bibfnamefont {P.}~\bibnamefont {K{\'o}m{\'a}r}}, \bibinfo {author} {\bibfnamefont {E.~M.}\ \bibnamefont {Kessler}}, \bibinfo {author} {\bibfnamefont {M.}~\bibnamefont {Bishof}}, \bibinfo {author} {\bibfnamefont {L.}~\bibnamefont {Jiang}}, \bibinfo {author} {\bibfnamefont {A.~S.}\ \bibnamefont {S{\o}rensen}}, \bibinfo {author} {\bibfnamefont {J.}~\bibnamefont {Ye}},\ and\ \bibinfo {author} {\bibfnamefont {M.~D.}\ \bibnamefont {Lukin}},\ }\bibfield  {title} {\bibinfo {title} {A quantum network of clocks},\ }\href {https://doi.org/10.1038/nphys3000} {\bibfield  {journal} {\bibinfo  {journal} {Nature Physics}\ }\textbf {\bibinfo {volume} {10}},\ \bibinfo {pages} {582} (\bibinfo {year} {2014})}\BibitemShut {NoStop}%
	\bibitem [{\citenamefont {Lu}\ \emph {et~al.}(2022)\citenamefont {Lu}, \citenamefont {Cao}, \citenamefont {Peng},\ and\ \citenamefont {Pan}}]{lu_micius_2022}%
	  \BibitemOpen
	  \bibfield  {author} {\bibinfo {author} {\bibfnamefont {C.-Y.}\ \bibnamefont {Lu}}, \bibinfo {author} {\bibfnamefont {Y.}~\bibnamefont {Cao}}, \bibinfo {author} {\bibfnamefont {C.-Z.}\ \bibnamefont {Peng}},\ and\ \bibinfo {author} {\bibfnamefont {J.-W.}\ \bibnamefont {Pan}},\ }\bibfield  {title} {\bibinfo {title} {Micius quantum experiments in space},\ }\href {https://doi.org/10.1103/RevModPhys.94.035001} {\bibfield  {journal} {\bibinfo  {journal} {Reviews of Modern Physics}\ }\textbf {\bibinfo {volume} {94}},\ \bibinfo {pages} {035001} (\bibinfo {year} {2022})}\BibitemShut {NoStop}%
	\bibitem [{\citenamefont {Inagaki}\ \emph {et~al.}(2013)\citenamefont {Inagaki}, \citenamefont {Matsuda}, \citenamefont {Tadanaga}, \citenamefont {Asobe},\ and\ \citenamefont {Takesue}}]{inagaki_entanglement_2013}%
	  \BibitemOpen
	  \bibfield  {author} {\bibinfo {author} {\bibfnamefont {T.}~\bibnamefont {Inagaki}}, \bibinfo {author} {\bibfnamefont {N.}~\bibnamefont {Matsuda}}, \bibinfo {author} {\bibfnamefont {O.}~\bibnamefont {Tadanaga}}, \bibinfo {author} {\bibfnamefont {M.}~\bibnamefont {Asobe}},\ and\ \bibinfo {author} {\bibfnamefont {H.}~\bibnamefont {Takesue}},\ }\bibfield  {title} {\bibinfo {title} {Entanglement distribution over 300 km of fiber},\ }\href {https://doi.org/10.1364/OE.21.023241} {\bibfield  {journal} {\bibinfo  {journal} {Optics Express}\ }\textbf {\bibinfo {volume} {21}},\ \bibinfo {pages} {23241} (\bibinfo {year} {2013})}\BibitemShut {NoStop}%
	\bibitem [{\citenamefont {Neumann}\ \emph {et~al.}(2022)\citenamefont {Neumann}, \citenamefont {Buchner}, \citenamefont {Bulla}, \citenamefont {Bohmann},\ and\ \citenamefont {Ursin}}]{neumann_entanglement_2022}%
	  \BibitemOpen
	  \bibfield  {author} {\bibinfo {author} {\bibfnamefont {S.~P.}\ \bibnamefont {Neumann}}, \bibinfo {author} {\bibfnamefont {A.}~\bibnamefont {Buchner}}, \bibinfo {author} {\bibfnamefont {L.}~\bibnamefont {Bulla}}, \bibinfo {author} {\bibfnamefont {M.}~\bibnamefont {Bohmann}},\ and\ \bibinfo {author} {\bibfnamefont {R.}~\bibnamefont {Ursin}},\ }\bibfield  {title} {\bibinfo {title} {Continuous entanglement distribution over a transnational 248 km fiber link},\ }\href {https://doi.org/10.1038/s41467-022-33919-0} {\bibfield  {journal} {\bibinfo  {journal} {Nature Communications}\ }\textbf {\bibinfo {volume} {13}},\ \bibinfo {pages} {6134} (\bibinfo {year} {2022})}\BibitemShut {NoStop}%
	\bibitem [{\citenamefont {Zhuang}\ \emph {et~al.}(2024)\citenamefont {Zhuang}, \citenamefont {Li}, \citenamefont {Zheng}, \citenamefont {Zeng}, \citenamefont {Wu}, \citenamefont {Li}, \citenamefont {Yao}, \citenamefont {Xie}, \citenamefont {Li}, \citenamefont {Qin}, \citenamefont {You}, \citenamefont {Xu}, \citenamefont {Yin}, \citenamefont {Cao}, \citenamefont {Zhang}, \citenamefont {Peng},\ and\ \citenamefont {Pan}}]{zhuang_ultrabright-entanglement-based_2024}%
	  \BibitemOpen
	  \bibfield  {author} {\bibinfo {author} {\bibfnamefont {S.-C.}\ \bibnamefont {Zhuang}}, \bibinfo {author} {\bibfnamefont {B.}~\bibnamefont {Li}}, \bibinfo {author} {\bibfnamefont {M.-Y.}\ \bibnamefont {Zheng}}, \bibinfo {author} {\bibfnamefont {Y.-X.}\ \bibnamefont {Zeng}}, \bibinfo {author} {\bibfnamefont {H.-N.}\ \bibnamefont {Wu}}, \bibinfo {author} {\bibfnamefont {G.-B.}\ \bibnamefont {Li}}, \bibinfo {author} {\bibfnamefont {Q.}~\bibnamefont {Yao}}, \bibinfo {author} {\bibfnamefont {X.-P.}\ \bibnamefont {Xie}}, \bibinfo {author} {\bibfnamefont {Y.-H.}\ \bibnamefont {Li}}, \bibinfo {author} {\bibfnamefont {H.}~\bibnamefont {Qin}}, \bibinfo {author} {\bibfnamefont {L.-X.}\ \bibnamefont {You}}, \bibinfo {author} {\bibfnamefont {F.-H.}\ \bibnamefont {Xu}}, \bibinfo {author} {\bibfnamefont {J.}~\bibnamefont {Yin}}, \bibinfo {author} {\bibfnamefont {Y.}~\bibnamefont {Cao}}, \bibinfo {author} {\bibfnamefont {Q.}~\bibnamefont {Zhang}}, \bibinfo {author} {\bibfnamefont {C.-Z.}\ \bibnamefont {Peng}},\ and\ \bibinfo {author} {\bibfnamefont {J.-W.}\ \bibnamefont {Pan}},\ }\href {https://doi.org/10.48550/arXiv.2408.04361} {\bibinfo {title} {Ultrabright-entanglement-based quantum key distribution over a 404-km-long optical fiber}} (\bibinfo {year} {2024}),\ \bibinfo {note} {arXiv:2408.04361 [quant-ph]}\BibitemShut {NoStop}%
	\bibitem [{\citenamefont {Azuma}\ \emph {et~al.}(2023)\citenamefont {Azuma}, \citenamefont {Economou}, \citenamefont {Elkouss}, \citenamefont {Hilaire}, \citenamefont {Jiang}, \citenamefont {Lo},\ and\ \citenamefont {Tzitrin}}]{azuma_quantum_2023}%
	  \BibitemOpen
	  \bibfield  {author} {\bibinfo {author} {\bibfnamefont {K.}~\bibnamefont {Azuma}}, \bibinfo {author} {\bibfnamefont {S.~E.}\ \bibnamefont {Economou}}, \bibinfo {author} {\bibfnamefont {D.}~\bibnamefont {Elkouss}}, \bibinfo {author} {\bibfnamefont {P.}~\bibnamefont {Hilaire}}, \bibinfo {author} {\bibfnamefont {L.}~\bibnamefont {Jiang}}, \bibinfo {author} {\bibfnamefont {H.-K.}\ \bibnamefont {Lo}},\ and\ \bibinfo {author} {\bibfnamefont {I.}~\bibnamefont {Tzitrin}},\ }\bibfield  {title} {\bibinfo {title} {Quantum repeaters: {From} quantum networks to the quantum internet},\ }\href {https://doi.org/10.1103/RevModPhys.95.045006} {\bibfield  {journal} {\bibinfo  {journal} {Reviews of Modern Physics}\ }\textbf {\bibinfo {volume} {95}},\ \bibinfo {pages} {045006} (\bibinfo {year} {2023})}\BibitemShut {NoStop}%
	\bibitem [{\citenamefont {Kumar}(1990)}]{kumar_quantum_1990}%
	  \BibitemOpen
	  \bibfield  {author} {\bibinfo {author} {\bibfnamefont {P.}~\bibnamefont {Kumar}},\ }\bibfield  {title} {\bibinfo {title} {Quantum frequency conversion},\ }\href {https://doi.org/10.1364/OL.15.001476} {\bibfield  {journal} {\bibinfo  {journal} {Optics Letters}\ }\textbf {\bibinfo {volume} {15}},\ \bibinfo {pages} {1476} (\bibinfo {year} {1990})}\BibitemShut {NoStop}%
	\bibitem [{\citenamefont {Yu}\ \emph {et~al.}(2020)\citenamefont {Yu}, \citenamefont {Ma}, \citenamefont {Luo}, \citenamefont {Jing}, \citenamefont {Sun}, \citenamefont {Fang}, \citenamefont {Yang}, \citenamefont {Liu}, \citenamefont {Zheng}, \citenamefont {Xie}, \citenamefont {Zhang}, \citenamefont {You}, \citenamefont {Wang}, \citenamefont {Chen}, \citenamefont {Zhang}, \citenamefont {Bao},\ and\ \citenamefont {Pan}}]{yu_entanglement_2020}%
	  \BibitemOpen
	  \bibfield  {author} {\bibinfo {author} {\bibfnamefont {Y.}~\bibnamefont {Yu}}, \bibinfo {author} {\bibfnamefont {F.}~\bibnamefont {Ma}}, \bibinfo {author} {\bibfnamefont {X.-Y.}\ \bibnamefont {Luo}}, \bibinfo {author} {\bibfnamefont {B.}~\bibnamefont {Jing}}, \bibinfo {author} {\bibfnamefont {P.-F.}\ \bibnamefont {Sun}}, \bibinfo {author} {\bibfnamefont {R.-Z.}\ \bibnamefont {Fang}}, \bibinfo {author} {\bibfnamefont {C.-W.}\ \bibnamefont {Yang}}, \bibinfo {author} {\bibfnamefont {H.}~\bibnamefont {Liu}}, \bibinfo {author} {\bibfnamefont {M.-Y.}\ \bibnamefont {Zheng}}, \bibinfo {author} {\bibfnamefont {X.-P.}\ \bibnamefont {Xie}}, \bibinfo {author} {\bibfnamefont {W.-J.}\ \bibnamefont {Zhang}}, \bibinfo {author} {\bibfnamefont {L.-X.}\ \bibnamefont {You}}, \bibinfo {author} {\bibfnamefont {Z.}~\bibnamefont {Wang}}, \bibinfo {author} {\bibfnamefont {T.-Y.}\ \bibnamefont {Chen}}, \bibinfo {author} {\bibfnamefont {Q.}~\bibnamefont {Zhang}}, \bibinfo {author} {\bibfnamefont {X.-H.}\ \bibnamefont {Bao}},\ and\ \bibinfo {author} {\bibfnamefont {J.-W.}\ \bibnamefont {Pan}},\ }\bibfield  {title} {\bibinfo {title} {Entanglement of two quantum memories via fibres over dozens of kilometres},\ }\href {https://doi.org/10.1038/s41586-020-1976-7} {\bibfield  {journal} {\bibinfo  {journal} {Nature}\ }\textbf {\bibinfo {volume} {578}},\ \bibinfo {pages} {240} (\bibinfo {year} {2020})}\BibitemShut {NoStop}%
	\bibitem [{\citenamefont {van Leent}\ \emph {et~al.}(2022)\citenamefont {van Leent}, \citenamefont {Bock}, \citenamefont {Fertig}, \citenamefont {Garthoff}, \citenamefont {Eppelt}, \citenamefont {Zhou}, \citenamefont {Malik}, \citenamefont {Seubert}, \citenamefont {Bauer}, \citenamefont {Rosenfeld}, \citenamefont {Zhang}, \citenamefont {Becher},\ and\ \citenamefont {Weinfurter}}]{van_leent_entangling_2022}%
	  \BibitemOpen
	  \bibfield  {author} {\bibinfo {author} {\bibfnamefont {T.}~\bibnamefont {van Leent}}, \bibinfo {author} {\bibfnamefont {M.}~\bibnamefont {Bock}}, \bibinfo {author} {\bibfnamefont {F.}~\bibnamefont {Fertig}}, \bibinfo {author} {\bibfnamefont {R.}~\bibnamefont {Garthoff}}, \bibinfo {author} {\bibfnamefont {S.}~\bibnamefont {Eppelt}}, \bibinfo {author} {\bibfnamefont {Y.}~\bibnamefont {Zhou}}, \bibinfo {author} {\bibfnamefont {P.}~\bibnamefont {Malik}}, \bibinfo {author} {\bibfnamefont {M.}~\bibnamefont {Seubert}}, \bibinfo {author} {\bibfnamefont {T.}~\bibnamefont {Bauer}}, \bibinfo {author} {\bibfnamefont {W.}~\bibnamefont {Rosenfeld}}, \bibinfo {author} {\bibfnamefont {W.}~\bibnamefont {Zhang}}, \bibinfo {author} {\bibfnamefont {C.}~\bibnamefont {Becher}},\ and\ \bibinfo {author} {\bibfnamefont {H.}~\bibnamefont {Weinfurter}},\ }\bibfield  {title} {\bibinfo {title} {Entangling single atoms over 33 km telecom fibre},\ }\href {https://doi.org/10.1038/s41586-022-04764-4} {\bibfield  {journal} {\bibinfo  {journal} {Nature}\ }\textbf {\bibinfo {volume} {607}},\ \bibinfo {pages} {69} (\bibinfo {year} {2022})}\BibitemShut {NoStop}%
	\bibitem [{\citenamefont {Knaut}\ \emph {et~al.}(2024)\citenamefont {Knaut}, \citenamefont {Suleymanzade}, \citenamefont {Wei}, \citenamefont {Assumpcao}, \citenamefont {Stas}, \citenamefont {Huan}, \citenamefont {Machielse}, \citenamefont {Knall}, \citenamefont {Sutula}, \citenamefont {Baranes}, \citenamefont {Sinclair}, \citenamefont {De-Eknamkul}, \citenamefont {Levonian}, \citenamefont {Bhaskar}, \citenamefont {Park}, \citenamefont {Lončar},\ and\ \citenamefont {Lukin}}]{knaut_entanglement_2024}%
	  \BibitemOpen
	  \bibfield  {author} {\bibinfo {author} {\bibfnamefont {C.~M.}\ \bibnamefont {Knaut}}, \bibinfo {author} {\bibfnamefont {A.}~\bibnamefont {Suleymanzade}}, \bibinfo {author} {\bibfnamefont {Y.-C.}\ \bibnamefont {Wei}}, \bibinfo {author} {\bibfnamefont {D.~R.}\ \bibnamefont {Assumpcao}}, \bibinfo {author} {\bibfnamefont {P.-J.}\ \bibnamefont {Stas}}, \bibinfo {author} {\bibfnamefont {Y.~Q.}\ \bibnamefont {Huan}}, \bibinfo {author} {\bibfnamefont {B.}~\bibnamefont {Machielse}}, \bibinfo {author} {\bibfnamefont {E.~N.}\ \bibnamefont {Knall}}, \bibinfo {author} {\bibfnamefont {M.}~\bibnamefont {Sutula}}, \bibinfo {author} {\bibfnamefont {G.}~\bibnamefont {Baranes}}, \bibinfo {author} {\bibfnamefont {N.}~\bibnamefont {Sinclair}}, \bibinfo {author} {\bibfnamefont {C.}~\bibnamefont {De-Eknamkul}}, \bibinfo {author} {\bibfnamefont {D.~S.}\ \bibnamefont {Levonian}}, \bibinfo {author} {\bibfnamefont {M.~K.}\ \bibnamefont {Bhaskar}}, \bibinfo {author} {\bibfnamefont {H.}~\bibnamefont {Park}}, \bibinfo {author} {\bibfnamefont {M.}~\bibnamefont {Lončar}},\ and\ \bibinfo {author} {\bibfnamefont {M.~D.}\ \bibnamefont {Lukin}},\ }\bibfield  {title} {\bibinfo {title} {Entanglement of nanophotonic quantum memory nodes in a telecom network},\ }\href {https://doi.org/10.1038/s41586-024-07252-z} {\bibfield  {journal} {\bibinfo  {journal} {Nature}\ }\textbf {\bibinfo {volume} {629}},\ \bibinfo {pages} {573} (\bibinfo {year} {2024})}\BibitemShut {NoStop}%
	\bibitem [{\citenamefont {Stolk}\ \emph {et~al.}(2024)\citenamefont {Stolk}, \citenamefont {Van Der~Enden}, \citenamefont {Slater}, \citenamefont {Te~Raa-Derckx}, \citenamefont {Botma}, \citenamefont {Van~Rantwijk}, \citenamefont {Biemond}, \citenamefont {Hagen}, \citenamefont {Herfst}, \citenamefont {Koek}, \citenamefont {Meskers}, \citenamefont {Vollmer}, \citenamefont {Van~Zwet}, \citenamefont {Markham}, \citenamefont {Edmonds}, \citenamefont {Geus}, \citenamefont {Elsen}, \citenamefont {Jungbluth}, \citenamefont {Haefner}, \citenamefont {Tresp}, \citenamefont {Stuhler}, \citenamefont {Ritter},\ and\ \citenamefont {Hanson}}]{stolk_metropolitan-scale_2024}%
	  \BibitemOpen
	  \bibfield  {author} {\bibinfo {author} {\bibfnamefont {A.~J.}\ \bibnamefont {Stolk}}, \bibinfo {author} {\bibfnamefont {K.~L.}\ \bibnamefont {Van Der~Enden}}, \bibinfo {author} {\bibfnamefont {M.-C.}\ \bibnamefont {Slater}}, \bibinfo {author} {\bibfnamefont {I.}~\bibnamefont {Te~Raa-Derckx}}, \bibinfo {author} {\bibfnamefont {P.}~\bibnamefont {Botma}}, \bibinfo {author} {\bibfnamefont {J.}~\bibnamefont {Van~Rantwijk}}, \bibinfo {author} {\bibfnamefont {J.~J.~B.}\ \bibnamefont {Biemond}}, \bibinfo {author} {\bibfnamefont {R.~A.~J.}\ \bibnamefont {Hagen}}, \bibinfo {author} {\bibfnamefont {R.~W.}\ \bibnamefont {Herfst}}, \bibinfo {author} {\bibfnamefont {W.~D.}\ \bibnamefont {Koek}}, \bibinfo {author} {\bibfnamefont {A.~J.~H.}\ \bibnamefont {Meskers}}, \bibinfo {author} {\bibfnamefont {R.}~\bibnamefont {Vollmer}}, \bibinfo {author} {\bibfnamefont {E.~J.}\ \bibnamefont {Van~Zwet}}, \bibinfo {author} {\bibfnamefont {M.}~\bibnamefont {Markham}}, \bibinfo {author} {\bibfnamefont {A.~M.}\ \bibnamefont {Edmonds}}, \bibinfo {author} {\bibfnamefont {J.~F.}\ \bibnamefont {Geus}}, \bibinfo {author} {\bibfnamefont {F.}~\bibnamefont {Elsen}}, \bibinfo {author} {\bibfnamefont {B.}~\bibnamefont {Jungbluth}}, \bibinfo {author} {\bibfnamefont {C.}~\bibnamefont {Haefner}}, \bibinfo {author} {\bibfnamefont {C.}~\bibnamefont {Tresp}}, \bibinfo {author} {\bibfnamefont {J.}~\bibnamefont {Stuhler}}, \bibinfo {author} {\bibfnamefont {S.}~\bibnamefont {Ritter}},\ and\ \bibinfo {author} {\bibfnamefont {R.}~\bibnamefont {Hanson}},\ }\bibfield  {title} {\bibinfo {title} {Metropolitan-scale heralded entanglement of solid-state qubits},\ }\href {https://doi.org/10.1126/sciadv.adp6442} {\bibfield  {journal} {\bibinfo  {journal} {Science Advances}\ }\textbf {\bibinfo {volume} {10}},\ \bibinfo {pages} {eadp6442} (\bibinfo {year} {2024})}\BibitemShut {NoStop}%
	\bibitem [{\citenamefont {Liu}\ \emph {et~al.}(2024)\citenamefont {Liu}, \citenamefont {Luo}, \citenamefont {Yu}, \citenamefont {Wang}, \citenamefont {Wang}, \citenamefont {Hu}, \citenamefont {Li}, \citenamefont {Zheng}, \citenamefont {Yao}, \citenamefont {Yan}, \citenamefont {Teng}, \citenamefont {Jiang}, \citenamefont {Liu}, \citenamefont {Xie}, \citenamefont {Zhang}, \citenamefont {Mao}, \citenamefont {Jiang}, \citenamefont {Zhang}, \citenamefont {Bao},\ and\ \citenamefont {Pan}}]{liu_creation_2024}%
	  \BibitemOpen
	  \bibfield  {author} {\bibinfo {author} {\bibfnamefont {J.-L.}\ \bibnamefont {Liu}}, \bibinfo {author} {\bibfnamefont {X.-Y.}\ \bibnamefont {Luo}}, \bibinfo {author} {\bibfnamefont {Y.}~\bibnamefont {Yu}}, \bibinfo {author} {\bibfnamefont {C.-Y.}\ \bibnamefont {Wang}}, \bibinfo {author} {\bibfnamefont {B.}~\bibnamefont {Wang}}, \bibinfo {author} {\bibfnamefont {Y.}~\bibnamefont {Hu}}, \bibinfo {author} {\bibfnamefont {J.}~\bibnamefont {Li}}, \bibinfo {author} {\bibfnamefont {M.-Y.}\ \bibnamefont {Zheng}}, \bibinfo {author} {\bibfnamefont {B.}~\bibnamefont {Yao}}, \bibinfo {author} {\bibfnamefont {Z.}~\bibnamefont {Yan}}, \bibinfo {author} {\bibfnamefont {D.}~\bibnamefont {Teng}}, \bibinfo {author} {\bibfnamefont {J.-W.}\ \bibnamefont {Jiang}}, \bibinfo {author} {\bibfnamefont {X.-B.}\ \bibnamefont {Liu}}, \bibinfo {author} {\bibfnamefont {X.-P.}\ \bibnamefont {Xie}}, \bibinfo {author} {\bibfnamefont {J.}~\bibnamefont {Zhang}}, \bibinfo {author} {\bibfnamefont {Q.-H.}\ \bibnamefont {Mao}}, \bibinfo {author} {\bibfnamefont {X.}~\bibnamefont {Jiang}}, \bibinfo {author} {\bibfnamefont {Q.}~\bibnamefont {Zhang}}, \bibinfo {author} {\bibfnamefont {X.-H.}\ \bibnamefont {Bao}},\ and\ \bibinfo {author} {\bibfnamefont {J.-W.}\ \bibnamefont {Pan}},\ }\bibfield  {title} {\bibinfo {title} {Creation of memory–memory entanglement in a metropolitan quantum network},\ }\href {https://doi.org/10.1038/s41586-024-07308-0} {\bibfield  {journal} {\bibinfo  {journal} {Nature}\ }\textbf {\bibinfo {volume} {629}},\ \bibinfo {pages} {579} (\bibinfo {year} {2024})}\BibitemShut {NoStop}%
	\bibitem [{\citenamefont {Duan}\ \emph {et~al.}(2001)\citenamefont {Duan}, \citenamefont {Lukin}, \citenamefont {Cirac},\ and\ \citenamefont {Zoller}}]{duan_long-distance_2001}%
	  \BibitemOpen
	  \bibfield  {author} {\bibinfo {author} {\bibfnamefont {L.-M.}\ \bibnamefont {Duan}}, \bibinfo {author} {\bibfnamefont {M.~D.}\ \bibnamefont {Lukin}}, \bibinfo {author} {\bibfnamefont {J.~I.}\ \bibnamefont {Cirac}},\ and\ \bibinfo {author} {\bibfnamefont {P.}~\bibnamefont {Zoller}},\ }\bibfield  {title} {\bibinfo {title} {Long-distance quantum communication with atomic ensembles and linear optics},\ }\href {https://doi.org/10.1038/35106500} {\bibfield  {journal} {\bibinfo  {journal} {Nature}\ }\textbf {\bibinfo {volume} {414}},\ \bibinfo {pages} {413} (\bibinfo {year} {2001})}\BibitemShut {NoStop}%
	\bibitem [{\citenamefont {Pirandola}\ \emph {et~al.}(2017)\citenamefont {Pirandola}, \citenamefont {Laurenza}, \citenamefont {Ottaviani},\ and\ \citenamefont {Banchi}}]{pirandola_fundamental_2017}%
	  \BibitemOpen
	  \bibfield  {author} {\bibinfo {author} {\bibfnamefont {S.}~\bibnamefont {Pirandola}}, \bibinfo {author} {\bibfnamefont {R.}~\bibnamefont {Laurenza}}, \bibinfo {author} {\bibfnamefont {C.}~\bibnamefont {Ottaviani}},\ and\ \bibinfo {author} {\bibfnamefont {L.}~\bibnamefont {Banchi}},\ }\bibfield  {title} {\bibinfo {title} {Fundamental limits of repeaterless quantum communications},\ }\href {https://doi.org/10.1038/ncomms15043} {\bibfield  {journal} {\bibinfo  {journal} {Nature Communications}\ }\textbf {\bibinfo {volume} {8}},\ \bibinfo {pages} {15043} (\bibinfo {year} {2017})}\BibitemShut {NoStop}%
	\bibitem [{\citenamefont {Yin}\ \emph {et~al.}(2017)\citenamefont {Yin}, \citenamefont {Cao}, \citenamefont {Li}, \citenamefont {Liao}, \citenamefont {Zhang}, \citenamefont {Ren}, \citenamefont {Cai}, \citenamefont {Liu}, \citenamefont {Li}, \citenamefont {Dai}, \citenamefont {Li}, \citenamefont {Lu}, \citenamefont {Gong}, \citenamefont {Xu}, \citenamefont {Li}, \citenamefont {Li}, \citenamefont {Yin}, \citenamefont {Jiang}, \citenamefont {Li}, \citenamefont {Jia}, \citenamefont {Ren}, \citenamefont {He}, \citenamefont {Zhou}, \citenamefont {Zhang}, \citenamefont {Wang}, \citenamefont {Chang}, \citenamefont {Zhu}, \citenamefont {Liu}, \citenamefont {Chen}, \citenamefont {Lu}, \citenamefont {Shu}, \citenamefont {Peng}, \citenamefont {Wang},\ and\ \citenamefont {Pan}}]{yin_satellite-based_2017}%
	  \BibitemOpen
	  \bibfield  {author} {\bibinfo {author} {\bibfnamefont {J.}~\bibnamefont {Yin}}, \bibinfo {author} {\bibfnamefont {Y.}~\bibnamefont {Cao}}, \bibinfo {author} {\bibfnamefont {Y.-H.}\ \bibnamefont {Li}}, \bibinfo {author} {\bibfnamefont {S.-K.}\ \bibnamefont {Liao}}, \bibinfo {author} {\bibfnamefont {L.}~\bibnamefont {Zhang}}, \bibinfo {author} {\bibfnamefont {J.-G.}\ \bibnamefont {Ren}}, \bibinfo {author} {\bibfnamefont {W.-Q.}\ \bibnamefont {Cai}}, \bibinfo {author} {\bibfnamefont {W.-Y.}\ \bibnamefont {Liu}}, \bibinfo {author} {\bibfnamefont {B.}~\bibnamefont {Li}}, \bibinfo {author} {\bibfnamefont {H.}~\bibnamefont {Dai}}, \bibinfo {author} {\bibfnamefont {G.-B.}\ \bibnamefont {Li}}, \bibinfo {author} {\bibfnamefont {Q.-M.}\ \bibnamefont {Lu}}, \bibinfo {author} {\bibfnamefont {Y.-H.}\ \bibnamefont {Gong}}, \bibinfo {author} {\bibfnamefont {Y.}~\bibnamefont {Xu}}, \bibinfo {author} {\bibfnamefont {S.-L.}\ \bibnamefont {Li}}, \bibinfo {author} {\bibfnamefont {F.-Z.}\ \bibnamefont {Li}}, \bibinfo {author} {\bibfnamefont {Y.-Y.}\ \bibnamefont {Yin}}, \bibinfo {author} {\bibfnamefont {Z.-Q.}\ \bibnamefont {Jiang}}, \bibinfo {author} {\bibfnamefont {M.}~\bibnamefont {Li}}, \bibinfo {author} {\bibfnamefont {J.-J.}\ \bibnamefont {Jia}}, \bibinfo {author} {\bibfnamefont {G.}~\bibnamefont {Ren}}, \bibinfo {author} {\bibfnamefont {D.}~\bibnamefont {He}}, \bibinfo {author} {\bibfnamefont {Y.-L.}\ \bibnamefont {Zhou}}, \bibinfo {author} {\bibfnamefont {X.-X.}\ \bibnamefont {Zhang}}, \bibinfo {author} {\bibfnamefont {N.}~\bibnamefont {Wang}}, \bibinfo {author} {\bibfnamefont {X.}~\bibnamefont {Chang}}, \bibinfo {author} {\bibfnamefont {Z.-C.}\ \bibnamefont {Zhu}}, \bibinfo {author} {\bibfnamefont {N.-L.}\ \bibnamefont {Liu}}, \bibinfo {author} {\bibfnamefont {Y.-A.}\ \bibnamefont {Chen}}, \bibinfo {author} {\bibfnamefont {C.-Y.}\ \bibnamefont {Lu}}, \bibinfo {author} {\bibfnamefont {R.}~\bibnamefont {Shu}}, \bibinfo {author} {\bibfnamefont {C.-Z.}\ \bibnamefont {Peng}}, \bibinfo {author} {\bibfnamefont {J.-Y.}\ \bibnamefont {Wang}},\ and\ \bibinfo {author} {\bibfnamefont {J.-W.}\ \bibnamefont {Pan}},\ }\bibfield  {title} {\bibinfo {title} {Satellite-based entanglement distribution over 1200 kilometers},\ }\href {https://doi.org/10.1126/science.aan3211} {\bibfield  {journal} {\bibinfo  {journal} {Science}\ }\textbf {\bibinfo {volume} {356}},\ \bibinfo {pages} {1140} (\bibinfo {year} {2017})}\BibitemShut {NoStop}%
	\bibitem [{\citenamefont {Chou}\ \emph {et~al.}(2005)\citenamefont {Chou}, \citenamefont {De~Riedmatten}, \citenamefont {Felinto}, \citenamefont {Polyakov}, \citenamefont {Van~Enk},\ and\ \citenamefont {Kimble}}]{chou_measurement-induced_2005}%
	  \BibitemOpen
	  \bibfield  {author} {\bibinfo {author} {\bibfnamefont {C.~W.}\ \bibnamefont {Chou}}, \bibinfo {author} {\bibfnamefont {H.}~\bibnamefont {De~Riedmatten}}, \bibinfo {author} {\bibfnamefont {D.}~\bibnamefont {Felinto}}, \bibinfo {author} {\bibfnamefont {S.~V.}\ \bibnamefont {Polyakov}}, \bibinfo {author} {\bibfnamefont {S.~J.}\ \bibnamefont {Van~Enk}},\ and\ \bibinfo {author} {\bibfnamefont {H.~J.}\ \bibnamefont {Kimble}},\ }\bibfield  {title} {\bibinfo {title} {Measurement-induced entanglement for excitation stored in remote atomic ensembles},\ }\href {https://doi.org/10.1038/nature04353} {\bibfield  {journal} {\bibinfo  {journal} {Nature}\ }\textbf {\bibinfo {volume} {438}},\ \bibinfo {pages} {828} (\bibinfo {year} {2005})}\BibitemShut {NoStop}%
	\bibitem [{\citenamefont {Chou}\ \emph {et~al.}(2007)\citenamefont {Chou}, \citenamefont {Laurat}, \citenamefont {Deng}, \citenamefont {Choi}, \citenamefont {de~Riedmatten}, \citenamefont {Felinto},\ and\ \citenamefont {Kimble}}]{chou_entanglement_2007}%
	  \BibitemOpen
	  \bibfield  {author} {\bibinfo {author} {\bibfnamefont {C.-W.}\ \bibnamefont {Chou}}, \bibinfo {author} {\bibfnamefont {J.}~\bibnamefont {Laurat}}, \bibinfo {author} {\bibfnamefont {H.}~\bibnamefont {Deng}}, \bibinfo {author} {\bibfnamefont {K.~S.}\ \bibnamefont {Choi}}, \bibinfo {author} {\bibfnamefont {H.}~\bibnamefont {de~Riedmatten}}, \bibinfo {author} {\bibfnamefont {D.}~\bibnamefont {Felinto}},\ and\ \bibinfo {author} {\bibfnamefont {H.~J.}\ \bibnamefont {Kimble}},\ }\bibfield  {title} {\bibinfo {title} {Functional quantum nodes for entanglement distribution over scalable quantum networks},\ }\href {https://doi.org/10.1126/science.1140300} {\bibfield  {journal} {\bibinfo  {journal} {Science}\ }\textbf {\bibinfo {volume} {316}},\ \bibinfo {pages} {1316} (\bibinfo {year} {2007})}\BibitemShut {NoStop}%
	\bibitem [{\citenamefont {Yuan}\ \emph {et~al.}(2008)\citenamefont {Yuan}, \citenamefont {Chen}, \citenamefont {Zhao}, \citenamefont {Chen}, \citenamefont {Schmiedmayer},\ and\ \citenamefont {Pan}}]{yuan_bdcz_2008}%
	  \BibitemOpen
	  \bibfield  {author} {\bibinfo {author} {\bibfnamefont {Z.-S.}\ \bibnamefont {Yuan}}, \bibinfo {author} {\bibfnamefont {Y.-A.}\ \bibnamefont {Chen}}, \bibinfo {author} {\bibfnamefont {B.}~\bibnamefont {Zhao}}, \bibinfo {author} {\bibfnamefont {S.}~\bibnamefont {Chen}}, \bibinfo {author} {\bibfnamefont {J.}~\bibnamefont {Schmiedmayer}},\ and\ \bibinfo {author} {\bibfnamefont {J.-W.}\ \bibnamefont {Pan}},\ }\bibfield  {title} {\bibinfo {title} {Experimental demonstration of a bdcz quantum repeater node},\ }\href {https://doi.org/10.1038/nature07241} {\bibfield  {journal} {\bibinfo  {journal} {Nature}\ }\textbf {\bibinfo {volume} {454}},\ \bibinfo {pages} {1098} (\bibinfo {year} {2008})}\BibitemShut {NoStop}%
	\bibitem [{\citenamefont {Lago-Rivera}\ \emph {et~al.}(2021)\citenamefont {Lago-Rivera}, \citenamefont {Grandi}, \citenamefont {Rakonjac}, \citenamefont {Seri},\ and\ \citenamefont {de~Riedmatten}}]{lago-rivera_rareearth_2021}%
	  \BibitemOpen
	  \bibfield  {author} {\bibinfo {author} {\bibfnamefont {D.}~\bibnamefont {Lago-Rivera}}, \bibinfo {author} {\bibfnamefont {S.}~\bibnamefont {Grandi}}, \bibinfo {author} {\bibfnamefont {J.~V.}\ \bibnamefont {Rakonjac}}, \bibinfo {author} {\bibfnamefont {A.}~\bibnamefont {Seri}},\ and\ \bibinfo {author} {\bibfnamefont {H.}~\bibnamefont {de~Riedmatten}},\ }\bibfield  {title} {\bibinfo {title} {Telecom-heralded entanglement between multimode solid-state quantum memories},\ }\href {https://doi.org/10.1038/s41586-021-03481-8} {\bibfield  {journal} {\bibinfo  {journal} {Nature}\ }\textbf {\bibinfo {volume} {594}},\ \bibinfo {pages} {37} (\bibinfo {year} {2021})}\BibitemShut {NoStop}%
	\bibitem [{\citenamefont {Liu}\ \emph {et~al.}(2021)\citenamefont {Liu}, \citenamefont {Hu}, \citenamefont {Li}, \citenamefont {Li}, \citenamefont {Li}, \citenamefont {Liang}, \citenamefont {Zhou}, \citenamefont {Li},\ and\ \citenamefont {Guo}}]{liu_rareearth_2021}%
	  \BibitemOpen
	  \bibfield  {author} {\bibinfo {author} {\bibfnamefont {X.}~\bibnamefont {Liu}}, \bibinfo {author} {\bibfnamefont {J.}~\bibnamefont {Hu}}, \bibinfo {author} {\bibfnamefont {Z.-F.}\ \bibnamefont {Li}}, \bibinfo {author} {\bibfnamefont {X.}~\bibnamefont {Li}}, \bibinfo {author} {\bibfnamefont {P.-Y.}\ \bibnamefont {Li}}, \bibinfo {author} {\bibfnamefont {P.-J.}\ \bibnamefont {Liang}}, \bibinfo {author} {\bibfnamefont {Z.-Q.}\ \bibnamefont {Zhou}}, \bibinfo {author} {\bibfnamefont {C.-F.}\ \bibnamefont {Li}},\ and\ \bibinfo {author} {\bibfnamefont {G.-C.}\ \bibnamefont {Guo}},\ }\bibfield  {title} {\bibinfo {title} {Heralded entanglement distribution between two absorptive quantum memories},\ }\href {https://doi.org/10.1038/s41586-021-03505-3} {\bibfield  {journal} {\bibinfo  {journal} {Nature}\ }\textbf {\bibinfo {volume} {594}},\ \bibinfo {pages} {41} (\bibinfo {year} {2021})}\BibitemShut {NoStop}%
	\bibitem [{\citenamefont {Delteil}\ \emph {et~al.}(2016)\citenamefont {Delteil}, \citenamefont {Sun}, \citenamefont {Gao}, \citenamefont {Togan}, \citenamefont {Faelt},\ and\ \citenamefont {Imamo{\u g}lu}}]{delteil_quantumdot_2016}%
	  \BibitemOpen
	  \bibfield  {author} {\bibinfo {author} {\bibfnamefont {A.}~\bibnamefont {Delteil}}, \bibinfo {author} {\bibfnamefont {Z.}~\bibnamefont {Sun}}, \bibinfo {author} {\bibfnamefont {W.-b.}\ \bibnamefont {Gao}}, \bibinfo {author} {\bibfnamefont {E.}~\bibnamefont {Togan}}, \bibinfo {author} {\bibfnamefont {S.}~\bibnamefont {Faelt}},\ and\ \bibinfo {author} {\bibfnamefont {A.}~\bibnamefont {Imamo{\u g}lu}},\ }\bibfield  {title} {\bibinfo {title} {Generation of heralded entanglement between distant hole spins},\ }\href {https://doi.org/10.1038/nphys3605} {\bibfield  {journal} {\bibinfo  {journal} {Nature Physics}\ }\textbf {\bibinfo {volume} {12}},\ \bibinfo {pages} {218} (\bibinfo {year} {2016})}\BibitemShut {NoStop}%
	\bibitem [{\citenamefont {Stockill}\ \emph {et~al.}(2017)\citenamefont {Stockill}, \citenamefont {Stanley}, \citenamefont {Huthmacher}, \citenamefont {Clarke}, \citenamefont {Hugues}, \citenamefont {Miller}, \citenamefont {Matthiesen}, \citenamefont {Le~Gall},\ and\ \citenamefont {Atat\"ure}}]{stockill_quantumdot_2017}%
	  \BibitemOpen
	  \bibfield  {author} {\bibinfo {author} {\bibfnamefont {R.}~\bibnamefont {Stockill}}, \bibinfo {author} {\bibfnamefont {M.~J.}\ \bibnamefont {Stanley}}, \bibinfo {author} {\bibfnamefont {L.}~\bibnamefont {Huthmacher}}, \bibinfo {author} {\bibfnamefont {E.}~\bibnamefont {Clarke}}, \bibinfo {author} {\bibfnamefont {M.}~\bibnamefont {Hugues}}, \bibinfo {author} {\bibfnamefont {A.~J.}\ \bibnamefont {Miller}}, \bibinfo {author} {\bibfnamefont {C.}~\bibnamefont {Matthiesen}}, \bibinfo {author} {\bibfnamefont {C.}~\bibnamefont {Le~Gall}},\ and\ \bibinfo {author} {\bibfnamefont {M.}~\bibnamefont {Atat\"ure}},\ }\bibfield  {title} {\bibinfo {title} {Phase-tuned entangled state generation between distant spin qubits},\ }\href {https://doi.org/10.1103/PhysRevLett.119.010503} {\bibfield  {journal} {\bibinfo  {journal} {Physical Review Letters}\ }\textbf {\bibinfo {volume} {119}},\ \bibinfo {pages} {010503} (\bibinfo {year} {2017})}\BibitemShut {NoStop}%
	\bibitem [{\citenamefont {Bernien}\ \emph {et~al.}(2013)\citenamefont {Bernien}, \citenamefont {Hensen}, \citenamefont {Pfaff}, \citenamefont {Koolstra}, \citenamefont {Blok}, \citenamefont {Robledo}, \citenamefont {Taminiau}, \citenamefont {Markham}, \citenamefont {Twitchen}, \citenamefont {Childress},\ and\ \citenamefont {Hanson}}]{bernien_colorcenter_2013}%
	  \BibitemOpen
	  \bibfield  {author} {\bibinfo {author} {\bibfnamefont {H.}~\bibnamefont {Bernien}}, \bibinfo {author} {\bibfnamefont {B.}~\bibnamefont {Hensen}}, \bibinfo {author} {\bibfnamefont {W.}~\bibnamefont {Pfaff}}, \bibinfo {author} {\bibfnamefont {G.}~\bibnamefont {Koolstra}}, \bibinfo {author} {\bibfnamefont {M.~S.}\ \bibnamefont {Blok}}, \bibinfo {author} {\bibfnamefont {L.}~\bibnamefont {Robledo}}, \bibinfo {author} {\bibfnamefont {T.~H.}\ \bibnamefont {Taminiau}}, \bibinfo {author} {\bibfnamefont {M.}~\bibnamefont {Markham}}, \bibinfo {author} {\bibfnamefont {D.~J.}\ \bibnamefont {Twitchen}}, \bibinfo {author} {\bibfnamefont {L.}~\bibnamefont {Childress}},\ and\ \bibinfo {author} {\bibfnamefont {R.}~\bibnamefont {Hanson}},\ }\bibfield  {title} {\bibinfo {title} {Heralded entanglement between solid-state qubits separated by three metres},\ }\href {https://doi.org/10.1038/nature12016} {\bibfield  {journal} {\bibinfo  {journal} {Nature}\ }\textbf {\bibinfo {volume} {497}},\ \bibinfo {pages} {86} (\bibinfo {year} {2013})}\BibitemShut {NoStop}%
	\bibitem [{\citenamefont {Hensen}\ \emph {et~al.}(2015)\citenamefont {Hensen}, \citenamefont {Bernien}, \citenamefont {Dr{\'e}au}, \citenamefont {Reiserer}, \citenamefont {Kalb}, \citenamefont {Blok}, \citenamefont {Ruitenberg}, \citenamefont {Vermeulen}, \citenamefont {Schouten}, \citenamefont {Abell{\'a}n}, \citenamefont {Amaya}, \citenamefont {Pruneri}, \citenamefont {Mitchell}, \citenamefont {Markham}, \citenamefont {Twitchen}, \citenamefont {Elkouss}, \citenamefont {Wehner}, \citenamefont {Taminiau},\ and\ \citenamefont {Hanson}}]{hensen_colorcenter_2015}%
	  \BibitemOpen
	  \bibfield  {author} {\bibinfo {author} {\bibfnamefont {B.}~\bibnamefont {Hensen}}, \bibinfo {author} {\bibfnamefont {H.}~\bibnamefont {Bernien}}, \bibinfo {author} {\bibfnamefont {A.~E.}\ \bibnamefont {Dr{\'e}au}}, \bibinfo {author} {\bibfnamefont {A.}~\bibnamefont {Reiserer}}, \bibinfo {author} {\bibfnamefont {N.}~\bibnamefont {Kalb}}, \bibinfo {author} {\bibfnamefont {M.~S.}\ \bibnamefont {Blok}}, \bibinfo {author} {\bibfnamefont {J.}~\bibnamefont {Ruitenberg}}, \bibinfo {author} {\bibfnamefont {R.~F.~L.}\ \bibnamefont {Vermeulen}}, \bibinfo {author} {\bibfnamefont {R.~N.}\ \bibnamefont {Schouten}}, \bibinfo {author} {\bibfnamefont {C.}~\bibnamefont {Abell{\'a}n}}, \bibinfo {author} {\bibfnamefont {W.}~\bibnamefont {Amaya}}, \bibinfo {author} {\bibfnamefont {V.}~\bibnamefont {Pruneri}}, \bibinfo {author} {\bibfnamefont {M.~W.}\ \bibnamefont {Mitchell}}, \bibinfo {author} {\bibfnamefont {M.}~\bibnamefont {Markham}}, \bibinfo {author} {\bibfnamefont {D.~J.}\ \bibnamefont {Twitchen}}, \bibinfo {author} {\bibfnamefont {D.}~\bibnamefont {Elkouss}}, \bibinfo {author} {\bibfnamefont {S.}~\bibnamefont {Wehner}}, \bibinfo {author} {\bibfnamefont {T.~H.}\ \bibnamefont {Taminiau}},\ and\ \bibinfo {author} {\bibfnamefont {R.}~\bibnamefont {Hanson}},\ }\bibfield  {title} {\bibinfo {title} {Loophole-free bell inequality violation using electron spins separated by 1.3 kilometres},\ }\href {https://doi.org/10.1038/nature15759} {\bibfield  {journal} {\bibinfo  {journal} {Nature}\ }\textbf {\bibinfo {volume} {526}},\ \bibinfo {pages} {682} (\bibinfo {year} {2015})}\BibitemShut {NoStop}%
	\bibitem [{\citenamefont {Humphreys}\ \emph {et~al.}(2018)\citenamefont {Humphreys}, \citenamefont {Kalb}, \citenamefont {Morits}, \citenamefont {Schouten}, \citenamefont {Vermeulen}, \citenamefont {Twitchen}, \citenamefont {Markham},\ and\ \citenamefont {Hanson}}]{humphreys_colorcenter_2018}%
	  \BibitemOpen
	  \bibfield  {author} {\bibinfo {author} {\bibfnamefont {P.~C.}\ \bibnamefont {Humphreys}}, \bibinfo {author} {\bibfnamefont {N.}~\bibnamefont {Kalb}}, \bibinfo {author} {\bibfnamefont {J.~P.~J.}\ \bibnamefont {Morits}}, \bibinfo {author} {\bibfnamefont {R.~N.}\ \bibnamefont {Schouten}}, \bibinfo {author} {\bibfnamefont {R.~F.~L.}\ \bibnamefont {Vermeulen}}, \bibinfo {author} {\bibfnamefont {D.~J.}\ \bibnamefont {Twitchen}}, \bibinfo {author} {\bibfnamefont {M.}~\bibnamefont {Markham}},\ and\ \bibinfo {author} {\bibfnamefont {R.}~\bibnamefont {Hanson}},\ }\bibfield  {title} {\bibinfo {title} {Deterministic delivery of remote entanglement on a quantum network},\ }\href {https://doi.org/10.1038/s41586-018-0200-5} {\bibfield  {journal} {\bibinfo  {journal} {Nature}\ }\textbf {\bibinfo {volume} {558}},\ \bibinfo {pages} {268} (\bibinfo {year} {2018})}\BibitemShut {NoStop}%
	\bibitem [{\citenamefont {Moehring}\ \emph {et~al.}(2007)\citenamefont {Moehring}, \citenamefont {Maunz}, \citenamefont {Olmschenk}, \citenamefont {Younge}, \citenamefont {Matsukevich}, \citenamefont {Duan},\ and\ \citenamefont {Monroe}}]{moehring_ions_2007}%
	  \BibitemOpen
	  \bibfield  {author} {\bibinfo {author} {\bibfnamefont {D.~L.}\ \bibnamefont {Moehring}}, \bibinfo {author} {\bibfnamefont {P.}~\bibnamefont {Maunz}}, \bibinfo {author} {\bibfnamefont {S.}~\bibnamefont {Olmschenk}}, \bibinfo {author} {\bibfnamefont {K.~C.}\ \bibnamefont {Younge}}, \bibinfo {author} {\bibfnamefont {D.~N.}\ \bibnamefont {Matsukevich}}, \bibinfo {author} {\bibfnamefont {L.~M.}\ \bibnamefont {Duan}},\ and\ \bibinfo {author} {\bibfnamefont {C.}~\bibnamefont {Monroe}},\ }\bibfield  {title} {\bibinfo {title} {Entanglement of single-atom quantum bits at a distance},\ }\href {https://doi.org/10.1038/nature06118} {\bibfield  {journal} {\bibinfo  {journal} {Nature}\ }\textbf {\bibinfo {volume} {449}},\ \bibinfo {pages} {68} (\bibinfo {year} {2007})}\BibitemShut {NoStop}%
	\bibitem [{\citenamefont {Hofmann}\ \emph {et~al.}(2012)\citenamefont {Hofmann}, \citenamefont {Krug}, \citenamefont {Ortegel}, \citenamefont {Gérard}, \citenamefont {Weber}, \citenamefont {Rosenfeld},\ and\ \citenamefont {Weinfurter}}]{hofmann_singleatom_2012}%
	  \BibitemOpen
	  \bibfield  {author} {\bibinfo {author} {\bibfnamefont {J.}~\bibnamefont {Hofmann}}, \bibinfo {author} {\bibfnamefont {M.}~\bibnamefont {Krug}}, \bibinfo {author} {\bibfnamefont {N.}~\bibnamefont {Ortegel}}, \bibinfo {author} {\bibfnamefont {L.}~\bibnamefont {Gérard}}, \bibinfo {author} {\bibfnamefont {M.}~\bibnamefont {Weber}}, \bibinfo {author} {\bibfnamefont {W.}~\bibnamefont {Rosenfeld}},\ and\ \bibinfo {author} {\bibfnamefont {H.}~\bibnamefont {Weinfurter}},\ }\bibfield  {title} {\bibinfo {title} {Heralded entanglement between widely separated atoms},\ }\href {https://doi.org/10.1126/science.1221856} {\bibfield  {journal} {\bibinfo  {journal} {Science}\ }\textbf {\bibinfo {volume} {337}},\ \bibinfo {pages} {72} (\bibinfo {year} {2012})}\BibitemShut {NoStop}%
	\bibitem [{\citenamefont {Zhou}\ \emph {et~al.}(2024)\citenamefont {Zhou}, \citenamefont {Malik}, \citenamefont {Fertig}, \citenamefont {Bock}, \citenamefont {Bauer}, \citenamefont {van Leent}, \citenamefont {Zhang}, \citenamefont {Becher},\ and\ \citenamefont {Weinfurter}}]{zhou_atomphotonentanglement_2024}%
	  \BibitemOpen
	  \bibfield  {author} {\bibinfo {author} {\bibfnamefont {Y.}~\bibnamefont {Zhou}}, \bibinfo {author} {\bibfnamefont {P.}~\bibnamefont {Malik}}, \bibinfo {author} {\bibfnamefont {F.}~\bibnamefont {Fertig}}, \bibinfo {author} {\bibfnamefont {M.}~\bibnamefont {Bock}}, \bibinfo {author} {\bibfnamefont {T.}~\bibnamefont {Bauer}}, \bibinfo {author} {\bibfnamefont {T.}~\bibnamefont {van Leent}}, \bibinfo {author} {\bibfnamefont {W.}~\bibnamefont {Zhang}}, \bibinfo {author} {\bibfnamefont {C.}~\bibnamefont {Becher}},\ and\ \bibinfo {author} {\bibfnamefont {H.}~\bibnamefont {Weinfurter}},\ }\bibfield  {title} {\bibinfo {title} {Long-lived quantum memory enabling atom-photon entanglement over 101 km of telecom fiber},\ }\href {https://doi.org/10.1103/PRXQuantum.5.020307} {\bibfield  {journal} {\bibinfo  {journal} {PRX Quantum}\ }\textbf {\bibinfo {volume} {5}},\ \bibinfo {pages} {020307} (\bibinfo {year} {2024})}\BibitemShut {NoStop}%
	\bibitem [{\citenamefont {Krutyanskiy}\ \emph {et~al.}(2024)\citenamefont {Krutyanskiy}, \citenamefont {Canteri}, \citenamefont {Meraner}, \citenamefont {Krcmarsky},\ and\ \citenamefont {Lanyon}}]{krutyanskiy_ionphotonentanglement_2024}%
	  \BibitemOpen
	  \bibfield  {author} {\bibinfo {author} {\bibfnamefont {V.}~\bibnamefont {Krutyanskiy}}, \bibinfo {author} {\bibfnamefont {M.}~\bibnamefont {Canteri}}, \bibinfo {author} {\bibfnamefont {M.}~\bibnamefont {Meraner}}, \bibinfo {author} {\bibfnamefont {V.}~\bibnamefont {Krcmarsky}},\ and\ \bibinfo {author} {\bibfnamefont {B.}~\bibnamefont {Lanyon}},\ }\bibfield  {title} {\bibinfo {title} {Multimode ion-photon entanglement over 101 kilometers},\ }\href {https://doi.org/10.1103/PRXQuantum.5.020308} {\bibfield  {journal} {\bibinfo  {journal} {PRX Quantum}\ }\textbf {\bibinfo {volume} {5}},\ \bibinfo {pages} {020308} (\bibinfo {year} {2024})}\BibitemShut {NoStop}%
	\bibitem [{\citenamefont {Simon}\ and\ \citenamefont {Irvine}(2003)}]{simon_entanglement_2003}%
	  \BibitemOpen
	  \bibfield  {author} {\bibinfo {author} {\bibfnamefont {C.}~\bibnamefont {Simon}}\ and\ \bibinfo {author} {\bibfnamefont {W.~T.~M.}\ \bibnamefont {Irvine}},\ }\bibfield  {title} {\bibinfo {title} {Robust long-distance entanglement and a loophole-free bell test with ions and photons},\ }\href {https://doi.org/10.1103/PhysRevLett.91.110405} {\bibfield  {journal} {\bibinfo  {journal} {Physical Review Letters}\ }\textbf {\bibinfo {volume} {91}},\ \bibinfo {pages} {110405} (\bibinfo {year} {2003})}\BibitemShut {NoStop}%
	\bibitem [{\citenamefont {Zhao}\ \emph {et~al.}(2007)\citenamefont {Zhao}, \citenamefont {Chen}, \citenamefont {Chen}, \citenamefont {Schmiedmayer},\ and\ \citenamefont {Pan}}]{zhao_entanglement_2007}%
	  \BibitemOpen
	  \bibfield  {author} {\bibinfo {author} {\bibfnamefont {B.}~\bibnamefont {Zhao}}, \bibinfo {author} {\bibfnamefont {Z.-B.}\ \bibnamefont {Chen}}, \bibinfo {author} {\bibfnamefont {Y.-A.}\ \bibnamefont {Chen}}, \bibinfo {author} {\bibfnamefont {J.}~\bibnamefont {Schmiedmayer}},\ and\ \bibinfo {author} {\bibfnamefont {J.-W.}\ \bibnamefont {Pan}},\ }\bibfield  {title} {\bibinfo {title} {Robust creation of entanglement between remote memory qubits},\ }\href {https://doi.org/10.1103/PhysRevLett.98.240502} {\bibfield  {journal} {\bibinfo  {journal} {Physical Review Letters}\ }\textbf {\bibinfo {volume} {98}},\ \bibinfo {pages} {240502} (\bibinfo {year} {2007})}\BibitemShut {NoStop}%
	\bibitem [{\citenamefont {Daiss}\ \emph {et~al.}(2021)\citenamefont {Daiss}, \citenamefont {Langenfeld}, \citenamefont {Welte}, \citenamefont {Distante}, \citenamefont {Thomas}, \citenamefont {Hartung}, \citenamefont {Morin},\ and\ \citenamefont {Rempe}}]{severin_gate_2021}%
	  \BibitemOpen
	  \bibfield  {author} {\bibinfo {author} {\bibfnamefont {S.}~\bibnamefont {Daiss}}, \bibinfo {author} {\bibfnamefont {S.}~\bibnamefont {Langenfeld}}, \bibinfo {author} {\bibfnamefont {S.}~\bibnamefont {Welte}}, \bibinfo {author} {\bibfnamefont {E.}~\bibnamefont {Distante}}, \bibinfo {author} {\bibfnamefont {P.}~\bibnamefont {Thomas}}, \bibinfo {author} {\bibfnamefont {L.}~\bibnamefont {Hartung}}, \bibinfo {author} {\bibfnamefont {O.}~\bibnamefont {Morin}},\ and\ \bibinfo {author} {\bibfnamefont {G.}~\bibnamefont {Rempe}},\ }\bibfield  {title} {\bibinfo {title} {A quantum-logic gate between distant quantum-network modules},\ }\href {https://doi.org/10.1126/science.abe3150} {\bibfield  {journal} {\bibinfo  {journal} {Science}\ }\textbf {\bibinfo {volume} {371}},\ \bibinfo {pages} {614} (\bibinfo {year} {2021})}\BibitemShut {NoStop}%
	\bibitem [{\citenamefont {Yang}\ \emph {et~al.}(2022)\citenamefont {Yang}, \citenamefont {Li}, \citenamefont {Zhou}, \citenamefont {Jiang}, \citenamefont {Bao},\ and\ \citenamefont {Pan}}]{yang_deterministic_2022}%
	  \BibitemOpen
	  \bibfield  {author} {\bibinfo {author} {\bibfnamefont {C.-W.}\ \bibnamefont {Yang}}, \bibinfo {author} {\bibfnamefont {J.}~\bibnamefont {Li}}, \bibinfo {author} {\bibfnamefont {M.-T.}\ \bibnamefont {Zhou}}, \bibinfo {author} {\bibfnamefont {X.}~\bibnamefont {Jiang}}, \bibinfo {author} {\bibfnamefont {X.-H.}\ \bibnamefont {Bao}},\ and\ \bibinfo {author} {\bibfnamefont {J.-W.}\ \bibnamefont {Pan}},\ }\bibfield  {title} {\bibinfo {title} {Deterministic measurement of a {Rydberg} superatom qubit via cavity-enhanced single-photon emission},\ }\href {https://doi.org/10.1364/OPTICA.461287} {\bibfield  {journal} {\bibinfo  {journal} {Optica}\ }\textbf {\bibinfo {volume} {9}},\ \bibinfo {pages} {853} (\bibinfo {year} {2022})}\BibitemShut {NoStop}%
	\bibitem [{\citenamefont {Zhang}\ \emph {et~al.}(2022)\citenamefont {Zhang}, \citenamefont {van Leent}, \citenamefont {Redeker}, \citenamefont {Garthoff}, \citenamefont {Schwonnek}, \citenamefont {Fertig}, \citenamefont {Eppelt}, \citenamefont {Rosenfeld}, \citenamefont {Scarani}, \citenamefont {Lim},\ and\ \citenamefont {Weinfurter}}]{zhang_device-independent_2022}%
	  \BibitemOpen
	  \bibfield  {author} {\bibinfo {author} {\bibfnamefont {W.}~\bibnamefont {Zhang}}, \bibinfo {author} {\bibfnamefont {T.}~\bibnamefont {van Leent}}, \bibinfo {author} {\bibfnamefont {K.}~\bibnamefont {Redeker}}, \bibinfo {author} {\bibfnamefont {R.}~\bibnamefont {Garthoff}}, \bibinfo {author} {\bibfnamefont {R.}~\bibnamefont {Schwonnek}}, \bibinfo {author} {\bibfnamefont {F.}~\bibnamefont {Fertig}}, \bibinfo {author} {\bibfnamefont {S.}~\bibnamefont {Eppelt}}, \bibinfo {author} {\bibfnamefont {W.}~\bibnamefont {Rosenfeld}}, \bibinfo {author} {\bibfnamefont {V.}~\bibnamefont {Scarani}}, \bibinfo {author} {\bibfnamefont {C.~C.-W.}\ \bibnamefont {Lim}},\ and\ \bibinfo {author} {\bibfnamefont {H.}~\bibnamefont {Weinfurter}},\ }\bibfield  {title} {\bibinfo {title} {A device-independent quantum key distribution system for distant users},\ }\href {https://doi.org/10.1038/s41586-022-04891-y} {\bibfield  {journal} {\bibinfo  {journal} {Nature}\ }\textbf {\bibinfo {volume} {607}},\ \bibinfo {pages} {687} (\bibinfo {year} {2022})}\BibitemShut {NoStop}%
	\bibitem [{\citenamefont {Nadlinger}\ \emph {et~al.}(2022)\citenamefont {Nadlinger}, \citenamefont {Drmota}, \citenamefont {Nichol}, \citenamefont {Araneda}, \citenamefont {Main}, \citenamefont {Srinivas}, \citenamefont {Lucas}, \citenamefont {Ballance}, \citenamefont {Ivanov}, \citenamefont {Tan}, \citenamefont {Sekatski}, \citenamefont {Urbanke}, \citenamefont {Renner}, \citenamefont {Sangouard},\ and\ \citenamefont {Bancal}}]{nadlinger_experimental_2022}%
	  \BibitemOpen
	  \bibfield  {author} {\bibinfo {author} {\bibfnamefont {D.~P.}\ \bibnamefont {Nadlinger}}, \bibinfo {author} {\bibfnamefont {P.}~\bibnamefont {Drmota}}, \bibinfo {author} {\bibfnamefont {B.~C.}\ \bibnamefont {Nichol}}, \bibinfo {author} {\bibfnamefont {G.}~\bibnamefont {Araneda}}, \bibinfo {author} {\bibfnamefont {D.}~\bibnamefont {Main}}, \bibinfo {author} {\bibfnamefont {R.}~\bibnamefont {Srinivas}}, \bibinfo {author} {\bibfnamefont {D.~M.}\ \bibnamefont {Lucas}}, \bibinfo {author} {\bibfnamefont {C.~J.}\ \bibnamefont {Ballance}}, \bibinfo {author} {\bibfnamefont {K.}~\bibnamefont {Ivanov}}, \bibinfo {author} {\bibfnamefont {E.~Y.-Z.}\ \bibnamefont {Tan}}, \bibinfo {author} {\bibfnamefont {P.}~\bibnamefont {Sekatski}}, \bibinfo {author} {\bibfnamefont {R.~L.}\ \bibnamefont {Urbanke}}, \bibinfo {author} {\bibfnamefont {R.}~\bibnamefont {Renner}}, \bibinfo {author} {\bibfnamefont {N.}~\bibnamefont {Sangouard}},\ and\ \bibinfo {author} {\bibfnamefont {J.-D.}\ \bibnamefont {Bancal}},\ }\bibfield  {title} {\bibinfo {title} {Experimental quantum key distribution certified by {Bell}'s theorem},\ }\href {https://doi.org/10.1038/s41586-022-04941-5} {\bibfield  {journal} {\bibinfo  {journal} {Nature}\ }\textbf {\bibinfo {volume} {607}},\ \bibinfo {pages} {682} (\bibinfo {year} {2022})}\BibitemShut {NoStop}%
	\bibitem [{\citenamefont {Yang}\ \emph {et~al.}(2016)\citenamefont {Yang}, \citenamefont {Wang}, \citenamefont {Bao},\ and\ \citenamefont {Pan}}]{yang_efficient_2016}%
	  \BibitemOpen
	  \bibfield  {author} {\bibinfo {author} {\bibfnamefont {S.-J.}\ \bibnamefont {Yang}}, \bibinfo {author} {\bibfnamefont {X.-J.}\ \bibnamefont {Wang}}, \bibinfo {author} {\bibfnamefont {X.-H.}\ \bibnamefont {Bao}},\ and\ \bibinfo {author} {\bibfnamefont {J.-W.}\ \bibnamefont {Pan}},\ }\bibfield  {title} {\bibinfo {title} {An efficient quantum light–matter interface with sub-second lifetime},\ }\href {https://doi.org/10.1038/nphoton.2016.51} {\bibfield  {journal} {\bibinfo  {journal} {Nature Photonics}\ }\textbf {\bibinfo {volume} {10}},\ \bibinfo {pages} {381} (\bibinfo {year} {2016})}\BibitemShut {NoStop}%
	\bibitem [{\citenamefont {Wang}\ \emph {et~al.}(2021)\citenamefont {Wang}, \citenamefont {Yang}, \citenamefont {Sun}, \citenamefont {Jing}, \citenamefont {Li}, \citenamefont {Zhou}, \citenamefont {Bao},\ and\ \citenamefont {Pan}}]{wang_cavity-enhanced_2021}%
	  \BibitemOpen
	  \bibfield  {author} {\bibinfo {author} {\bibfnamefont {X.-J.}\ \bibnamefont {Wang}}, \bibinfo {author} {\bibfnamefont {S.-J.}\ \bibnamefont {Yang}}, \bibinfo {author} {\bibfnamefont {P.-F.}\ \bibnamefont {Sun}}, \bibinfo {author} {\bibfnamefont {B.}~\bibnamefont {Jing}}, \bibinfo {author} {\bibfnamefont {J.}~\bibnamefont {Li}}, \bibinfo {author} {\bibfnamefont {M.-T.}\ \bibnamefont {Zhou}}, \bibinfo {author} {\bibfnamefont {X.-H.}\ \bibnamefont {Bao}},\ and\ \bibinfo {author} {\bibfnamefont {J.-W.}\ \bibnamefont {Pan}},\ }\bibfield  {title} {\bibinfo {title} {Cavity-{Enhanced} {Atom}-{Photon} {Entanglement} with {Subsecond} {Lifetime}},\ }\href {https://doi.org/10.1103/PhysRevLett.126.090501} {\bibfield  {journal} {\bibinfo  {journal} {Physical Review Letters}\ }\textbf {\bibinfo {volume} {126}},\ \bibinfo {pages} {090501} (\bibinfo {year} {2021})}\BibitemShut {NoStop}%
	\bibitem [{\citenamefont {Luo}\ \emph {et~al.}(2022)\citenamefont {Luo}, \citenamefont {Yu}, \citenamefont {Liu}, \citenamefont {Zheng}, \citenamefont {Wang}, \citenamefont {Wang}, \citenamefont {Li}, \citenamefont {Jiang}, \citenamefont {Xie}, \citenamefont {Zhang}, \citenamefont {Bao},\ and\ \citenamefont {Pan}}]{luo_postselected_2022}%
	  \BibitemOpen
	  \bibfield  {author} {\bibinfo {author} {\bibfnamefont {X.-Y.}\ \bibnamefont {Luo}}, \bibinfo {author} {\bibfnamefont {Y.}~\bibnamefont {Yu}}, \bibinfo {author} {\bibfnamefont {J.-L.}\ \bibnamefont {Liu}}, \bibinfo {author} {\bibfnamefont {M.-Y.}\ \bibnamefont {Zheng}}, \bibinfo {author} {\bibfnamefont {C.-Y.}\ \bibnamefont {Wang}}, \bibinfo {author} {\bibfnamefont {B.}~\bibnamefont {Wang}}, \bibinfo {author} {\bibfnamefont {J.}~\bibnamefont {Li}}, \bibinfo {author} {\bibfnamefont {X.}~\bibnamefont {Jiang}}, \bibinfo {author} {\bibfnamefont {X.-P.}\ \bibnamefont {Xie}}, \bibinfo {author} {\bibfnamefont {Q.}~\bibnamefont {Zhang}}, \bibinfo {author} {\bibfnamefont {X.-H.}\ \bibnamefont {Bao}},\ and\ \bibinfo {author} {\bibfnamefont {J.-W.}\ \bibnamefont {Pan}},\ }\bibfield  {title} {\bibinfo {title} {Postselected {Entanglement} between {Two} {Atomic} {Ensembles} {Separated} by 12.5 km},\ }\href {https://doi.org/10.1103/PhysRevLett.129.050503} {\bibfield  {journal} {\bibinfo  {journal} {Physical Review Letters}\ }\textbf {\bibinfo {volume} {129}},\ \bibinfo {pages} {050503} (\bibinfo {year} {2022})}\BibitemShut {NoStop}%
	\bibitem [{\citenamefont {Zaske}\ \emph {et~al.}(2011)\citenamefont {Zaske}, \citenamefont {Lenhard},\ and\ \citenamefont {Becher}}]{zaske_efficient_2011}%
	  \BibitemOpen
	  \bibfield  {author} {\bibinfo {author} {\bibfnamefont {S.}~\bibnamefont {Zaske}}, \bibinfo {author} {\bibfnamefont {A.}~\bibnamefont {Lenhard}},\ and\ \bibinfo {author} {\bibfnamefont {C.}~\bibnamefont {Becher}},\ }\bibfield  {title} {\bibinfo {title} {Efficient frequency downconversion at the single photon level from the red spectral range to the telecommunications {C}-band},\ }\href {https://doi.org/10.1364/OE.19.012825} {\bibfield  {journal} {\bibinfo  {journal} {Optics Express}\ }\textbf {\bibinfo {volume} {19}},\ \bibinfo {pages} {12825} (\bibinfo {year} {2011})}\BibitemShut {NoStop}%
	\bibitem [{\citenamefont {Van~Leent}\ \emph {et~al.}(2020)\citenamefont {Van~Leent}, \citenamefont {Bock}, \citenamefont {Garthoff}, \citenamefont {Redeker}, \citenamefont {Zhang}, \citenamefont {Bauer}, \citenamefont {Rosenfeld}, \citenamefont {Becher},\ and\ \citenamefont {Weinfurter}}]{van_leent_long-distance_2020}%
	  \BibitemOpen
	  \bibfield  {author} {\bibinfo {author} {\bibfnamefont {T.}~\bibnamefont {Van~Leent}}, \bibinfo {author} {\bibfnamefont {M.}~\bibnamefont {Bock}}, \bibinfo {author} {\bibfnamefont {R.}~\bibnamefont {Garthoff}}, \bibinfo {author} {\bibfnamefont {K.}~\bibnamefont {Redeker}}, \bibinfo {author} {\bibfnamefont {W.}~\bibnamefont {Zhang}}, \bibinfo {author} {\bibfnamefont {T.}~\bibnamefont {Bauer}}, \bibinfo {author} {\bibfnamefont {W.}~\bibnamefont {Rosenfeld}}, \bibinfo {author} {\bibfnamefont {C.}~\bibnamefont {Becher}},\ and\ \bibinfo {author} {\bibfnamefont {H.}~\bibnamefont {Weinfurter}},\ }\bibfield  {title} {\bibinfo {title} {Long-{Distance} {Distribution} of {Atom}-{Photon} {Entanglement} at {Telecom} {Wavelength}},\ }\href {https://doi.org/10.1103/PhysRevLett.124.010510} {\bibfield  {journal} {\bibinfo  {journal} {Physical Review Letters}\ }\textbf {\bibinfo {volume} {124}},\ \bibinfo {pages} {010510} (\bibinfo {year} {2020})}\BibitemShut {NoStop}%
	\bibitem [{\citenamefont {Ma}\ \emph {et~al.}(2012)\citenamefont {Ma}, \citenamefont {Zotter}, \citenamefont {Kofler}, \citenamefont {Ursin}, \citenamefont {Jennewein}, \citenamefont {Brukner},\ and\ \citenamefont {Zeilinger}}]{ma_experimental_2012}%
	  \BibitemOpen
	  \bibfield  {author} {\bibinfo {author} {\bibfnamefont {X.-S.}\ \bibnamefont {Ma}}, \bibinfo {author} {\bibfnamefont {S.}~\bibnamefont {Zotter}}, \bibinfo {author} {\bibfnamefont {J.}~\bibnamefont {Kofler}}, \bibinfo {author} {\bibfnamefont {R.}~\bibnamefont {Ursin}}, \bibinfo {author} {\bibfnamefont {T.}~\bibnamefont {Jennewein}}, \bibinfo {author} {\bibfnamefont {{\v C}.}~\bibnamefont {Brukner}},\ and\ \bibinfo {author} {\bibfnamefont {A.}~\bibnamefont {Zeilinger}},\ }\bibfield  {title} {\bibinfo {title} {Experimental delayed-choice entanglement swapping},\ }\href {https://doi.org/10.1038/nphys2294} {\bibfield  {journal} {\bibinfo  {journal} {Nature Physics}\ }\textbf {\bibinfo {volume} {8}},\ \bibinfo {pages} {479} (\bibinfo {year} {2012})}\BibitemShut {NoStop}%
	\bibitem [{\citenamefont {Lucamarini}\ \emph {et~al.}(2018)\citenamefont {Lucamarini}, \citenamefont {Yuan}, \citenamefont {Dynes},\ and\ \citenamefont {Shields}}]{lucamarini_overcoming_2018}%
	  \BibitemOpen
	  \bibfield  {author} {\bibinfo {author} {\bibfnamefont {M.}~\bibnamefont {Lucamarini}}, \bibinfo {author} {\bibfnamefont {Z.~L.}\ \bibnamefont {Yuan}}, \bibinfo {author} {\bibfnamefont {J.~F.}\ \bibnamefont {Dynes}},\ and\ \bibinfo {author} {\bibfnamefont {A.~J.}\ \bibnamefont {Shields}},\ }\bibfield  {title} {\bibinfo {title} {Overcoming the rate-distance limit of quantum key distribution without quantum repeaters},\ }\href {https://doi.org/10.1038/s41586-018-0066-6} {\bibfield  {journal} {\bibinfo  {journal} {Nature}\ ,\ \bibinfo {pages} {1}} (\bibinfo {year} {2018})}\BibitemShut {NoStop}%
	\bibitem [{\citenamefont {Li}\ \emph {et~al.}(2013)\citenamefont {Li}, \citenamefont {Dudin},\ and\ \citenamefont {Kuzmich}}]{li_entanglement_2013}%
	  \BibitemOpen
	  \bibfield  {author} {\bibinfo {author} {\bibfnamefont {L.}~\bibnamefont {Li}}, \bibinfo {author} {\bibfnamefont {Y.~O.}\ \bibnamefont {Dudin}},\ and\ \bibinfo {author} {\bibfnamefont {A.}~\bibnamefont {Kuzmich}},\ }\bibfield  {title} {\bibinfo {title} {Entanglement between light and an optical atomic excitation},\ }\href {https://doi.org/10.1038/nature12227} {\bibfield  {journal} {\bibinfo  {journal} {Nature}\ }\textbf {\bibinfo {volume} {498}},\ \bibinfo {pages} {466} (\bibinfo {year} {2013})}\BibitemShut {NoStop}%
	\bibitem [{\citenamefont {Sun}\ \emph {et~al.}(2022)\citenamefont {Sun}, \citenamefont {Yu}, \citenamefont {An}, \citenamefont {Li}, \citenamefont {Yang}, \citenamefont {Bao},\ and\ \citenamefont {Pan}}]{sun_deterministic_2022}%
	  \BibitemOpen
	  \bibfield  {author} {\bibinfo {author} {\bibfnamefont {P.-F.}\ \bibnamefont {Sun}}, \bibinfo {author} {\bibfnamefont {Y.}~\bibnamefont {Yu}}, \bibinfo {author} {\bibfnamefont {Z.-Y.}\ \bibnamefont {An}}, \bibinfo {author} {\bibfnamefont {J.}~\bibnamefont {Li}}, \bibinfo {author} {\bibfnamefont {C.-W.}\ \bibnamefont {Yang}}, \bibinfo {author} {\bibfnamefont {X.-H.}\ \bibnamefont {Bao}},\ and\ \bibinfo {author} {\bibfnamefont {J.-W.}\ \bibnamefont {Pan}},\ }\bibfield  {title} {\bibinfo {title} {Deterministic {Time}-{Bin} {Entanglement} between a {Single} {Photon} and an {Atomic} {Ensemble}},\ }\href {https://doi.org/10.1103/PhysRevLett.128.060502} {\bibfield  {journal} {\bibinfo  {journal} {Physical Review Letters}\ }\textbf {\bibinfo {volume} {128}},\ \bibinfo {pages} {060502} (\bibinfo {year} {2022})}\BibitemShut {NoStop}%
	\bibitem [{\citenamefont {Briegel}\ \emph {et~al.}(1998)\citenamefont {Briegel}, \citenamefont {D\"ur}, \citenamefont {Cirac},\ and\ \citenamefont {Zoller}}]{briegel_repeater_1998}%
	  \BibitemOpen
	  \bibfield  {author} {\bibinfo {author} {\bibfnamefont {H.-J.}\ \bibnamefont {Briegel}}, \bibinfo {author} {\bibfnamefont {W.}~\bibnamefont {D\"ur}}, \bibinfo {author} {\bibfnamefont {J.~I.}\ \bibnamefont {Cirac}},\ and\ \bibinfo {author} {\bibfnamefont {P.}~\bibnamefont {Zoller}},\ }\bibfield  {title} {\bibinfo {title} {Quantum repeaters: The role of imperfect local operations in quantum communication},\ }\href {https://doi.org/10.1103/PhysRevLett.81.5932} {\bibfield  {journal} {\bibinfo  {journal} {Physical Review Letters}\ }\textbf {\bibinfo {volume} {81}},\ \bibinfo {pages} {5932} (\bibinfo {year} {1998})}\BibitemShut {NoStop}%
	\bibitem [{\citenamefont {Sangouard}\ \emph {et~al.}(2011)\citenamefont {Sangouard}, \citenamefont {Simon}, \citenamefont {de~Riedmatten},\ and\ \citenamefont {Gisin}}]{sangouard_quantum_2011}%
	  \BibitemOpen
	  \bibfield  {author} {\bibinfo {author} {\bibfnamefont {N.}~\bibnamefont {Sangouard}}, \bibinfo {author} {\bibfnamefont {C.}~\bibnamefont {Simon}}, \bibinfo {author} {\bibfnamefont {H.}~\bibnamefont {de~Riedmatten}},\ and\ \bibinfo {author} {\bibfnamefont {N.}~\bibnamefont {Gisin}},\ }\bibfield  {title} {\bibinfo {title} {Quantum repeaters based on atomic ensembles and linear optics},\ }\href {https://doi.org/10.1103/RevModPhys.83.33} {\bibfield  {journal} {\bibinfo  {journal} {Reviews of Modern Physics}\ }\textbf {\bibinfo {volume} {83}},\ \bibinfo {pages} {33} (\bibinfo {year} {2011})}\BibitemShut {NoStop}%
	\end{thebibliography}

\begin{thebibliography}{1}

	\bibitem{yu2020}
	Y. Yu, {\it et~al.\/}, {\it Nature\/} {\bf 578}, 240
	  (2020).
	
	\bibitem{zaske2011}
	S. Zask, {\it et~al.\/}, {\it Optics Express\/} {\bf 19},
	   12825 (2011).
	
	\bibitem{vanleent2020}
	T. van Leent, {\it et~al.\/}, {\it Physical Review Letters\/} {\bf 124}, 010510 (2020).
	
	\bibitem{tsujimoto2021}
	Y. Tsujimoto, {\it et~al.\/}, {\it Optics Express\/} {\bf 29}, 37150 (2021).
	
	\bibitem{liu2024}
	J.-L. Liu, {\it et~al.\/}, {\it
	  Nature\/} {\bf 629}, 579 (2024).
	
\end{thebibliography}

\end{document}